\documentclass[journal]{IEEEtran}
\usepackage[cmex10]{amsmath}
\usepackage{graphicx}
\usepackage{cite}
\usepackage{epstopdf}
\usepackage{algorithm}
\usepackage{algorithmic}
\usepackage[caption = false]{subfig}
\usepackage{booktabs}
\usepackage{amsfonts}
\usepackage{color}

\newtheorem{theorem}{\textbf{Theorem}}

\begin{document}
\title{Energy-Efficient Transmission of Hybrid Array with \mbox{Non-Ideal} Power Amplifiers and Circuitry}
\author{Yuhao~Zhang, Qimei~Cui,~\IEEEmembership{Senior~Member,~IEEE}, Wei~Ni,~\IEEEmembership{Senior~Member,~IEEE}, and~Ping~Zhang
\IEEEcompsocitemizethanks{
\IEEEcompsocthanksitem The work was supported in part by the National Nature Science Foundation of China Project under Grant 61471058, in part by the Hong Kong, Macao and Taiwan Science and Technology Cooperation Projects under Grant 2016YFE0122900, in part by the Beijing Science and Technology Commission Foundation under Grant 201702005, and in part by the 111 Project of China under Grant B16006. \textit{(Corresponding author: Qimei Cui.)}
\IEEEcompsocthanksitem Y. Zhang, Q. Cui, and P. Zhang are with the National Engineering Laboratory for Mobile Network Technologies, Beijing University of Posts and Telecommunications, Beijing, 100876, China (e-mail: cuiqimei@bupt.edu.cn).

W. Ni is with Commonwealth Scientific and Industrial Research Organisation (CSIRO), Marsfield, Sydney, NSW 2122, Australia (e-mail: wei.ni@csiro.au).}}

\maketitle


\begin{abstract}
This paper presents a new approach to efficiently maximizing the energy efficiency~(EE) of hybrid arrays under a practical setting of \mbox{non-ideal} power amplifiers~(PAs) and \mbox{non-negligible} circuit power, where coherent and \mbox{non-coherent} beamforming are considered. As a key contribution, we reveal that a bursty transmission mode can be \mbox{energy-efficient} to achieve steady transmissions of a data stream under the practical setting. This is distinctively different from existing studies under ideal circuits and PAs, where continuous transmissions are the most energy-efficient. Another important contribution is that the optimal transmit duration and powers are identified to balance energy consumptions in the \mbox{non-ideal} circuits and PAs, and maximize the EE. This is achieved by establishing the most energy-efficient structure of transmit powers, given a transmit duration, and correspondingly partitioning the \mbox{non-convex} feasible region of the transmit duration into segments with \mbox{self-contained} convexity or concavity. Evident from simulations, significant EE gains of the proposed approach are demonstrated through comparisons with the state of the art, and the superiority of the bursty transmission mode is confirmed especially under low data rate demands.
\end{abstract}


\section{Introduction}\label{sec:1}
Equipped with tens to hundreds of antennas, massive MIMO is one of the promising technologies for improving spectral efficiency~(SE) and saving \mbox{per-antenna} transmit power~\cite{Ref3,Ref5,Ref2}. It is of particular importance to millimeter-Wave~(mmWave) communications by exploiting diversity and beamforming~(BF) to compensate for poor channel propagation~\cite{Ref8,Ref4}. Massive MIMO is also well suited for mmWave applications, due to significantly small antenna sizes in mmWave~\cite{Ref11,Ref15,Ref25}. However, with the increasing number of antennas, the total power consumption and implementation complexity of massive MIMO would increase. Hybrid arrays have been accepted as a practical implementation of massive MIMO, where a large-scale antenna array divides into an adequate number of analog phased subarrays. Digital processing is carried out upon the input and output of the subarrays~\cite{Ref17,Ref7,Ref12}. By this means, the requirement of accommodating large amounts of radio frequency~(RF) hardware, such as analog-to-digital/digital-to-analog converter~(ADC/DAC), in confined space can be relieved, and so can the complexity and energy requirements of array processing.

Hybrid arrays have been demonstrated to achieve high energy efficiency~(EE)~\cite{Ref16,Ref17,Ref7,Ref12}, which is a key performance index of the networks, and is critical to massive MIMO due to the use of large numbers of power amplifiers~(PAs) and RF circuits. This is because PAs and circuits can dramatically consume energy and penalize the EE of massive MIMO~\cite{Ref13}, especially in the practical case where the PAs are \mbox{non-ideal}. A substantial part of the power input to a PA is not used for data transmission, deteriorating the EE~\cite{circuitp1,circuitp2,Ref19,EEmetric}. Moreover, \mbox{non-ideal} PAs can render the optimization of the EE intractable, as the response function of \mbox{non-ideal} PAs is \mbox{non-linear} and \mbox{non-convex}~\cite{Ref43,Ref18}.

On the other hand, the \mbox{non-negligible} power consumption of transmitter circuits can also pose difficulties to maximizing the EE of large-scale antenna arrays. It necessitates new variable of transmit duration to be optimized, apart from the transmit powers of analog subarrays. Particularly, if a hybrid array transmits excessively long, the circuit energy consumption would increase and reduce the EE. On the other hand, if the array transmits too short, the transmit power becomes excessively high, which, in the presence of non-ideal PAs, is detrimental to the EE. Moreover, the \mbox{non-negligible} circuit power can be \mbox{non-linear} and \mbox{non-convex} to the transmit power, since it is typically linear to the data rate and hence logarithmic to the transmit power~\cite{EEmetric,Ref43,Ref18,EnergyCom}.

This paper presents a new approach to efficiently optimize transmit powers and duration for maximizing the EE of hybrid arrays in the presence of practical \mbox{non-ideal} PAs and \mbox{non-negligible} circuit power. Coherent and \mbox{non-coherent} beamforming techniques are considered under different availability of channel state information~(CSI). By decoupling the optimization of the transmit powers from that of the transmit duration, we discover the most energy-efficient structure of the transmit powers, given a transmit duration. The structure, in turn, can be used to partition the \mbox{non-convex} feasible solution region of the transmit duration into segments with \mbox{self-contained} convexity or concavity. Particularly, we prove that the EE is convex in one segment and concave in the rest under \mbox{non-coherent} beamforming, and is convex in all segments under coherent beamforming. The optimal transmit duration can therefore be efficiently solved by evaluating the boundaries and stationary points of the segments. Extensive simulations confirm our discovery and the superiority of our approach to the state of the art in terms of EE.

Another important contribution is that we reveal a bursty transmission mode can be more energy-efficient for a data stream with a consistent average rate requirement than continuous transmissions in the presence of \mbox{non-negligible} circuit power and \mbox{non-ideal} PAs. This is due to the fact that the hybrid array can be turned on only for part of a timeslot and remain off for the rest of the slot, so as to achieve the average data rate while reducing the circuit energy consumption associated with transmission. Part of non-negligible circuit energy consumption, such as those on ADC and up-converter, can be increasingly saved with the decrease of the transmit time. This is distinctively different from existing studies under ideal circuit and PAs, where continuous transmissions are the most energy-efficient due to the fact that the data rate is increased by either linearly increasing the transmit duration or exponentially increasing the transmit power under ideal circuit and/or PAs.

The rest of this paper is organized as follows. In Section~\ref{SecRelatedWork}, the related work is provided. In Section~\ref{sec:2}, the system model is described. In Section~\ref{sec:3}, the optimization problem is formulated. The structure of the optimal transmit powers is discovered in Section~\ref{sec:4}, based on which the feasible region of the transmit duration is segmented and optimized in Section~\ref{sec:5}. Simulation results are shown in Section~\ref{sec:7}, followed by conclusions in Section~\ref{sec:8}.
\section{Related Work}\label{SecRelatedWork}
The two state-of-the-art designs of hybrid arrays, namely, localized and interleaved hybrid arrays, were presented in~\cite{Ref7}. In~\cite{Ref17}, considering two different structures where the signal from each RF chain can be delivered to all antennas and to limited antennas respectively, two types of hybrid architecture were proposed in the multi-user scenario, based on which the trade-off of EE and SE is analyzed. As for the EE research, most of the existing works are conducted for the precoding and beamforming method. For example, in~\cite{Ref10}, the energy-efficient design of the precoder in hybrid array was investigated, and the non-convex EE optimization problem was solved by a two-layer optimization method, where the analog and digital precoders are optimized in an alternating manner. The convergence of the proposed scheme was proved by the monotonic boundary theorem and fractional programming theory. In~\cite{Ref14}, hybrid analog-digital transceivers were designed with fully and partially connected architectures to maximize the EE by deriving the precoding and combining matrices through decoupled \mbox{non-convex} transmitter-receiver optimization. Moreover, the EE performance was also examined w.r.t the number of RF chains and antennas.

Existing works on hybrid arrays have been extensively based on the assumption of ideal circuitry and PAs~\cite{Ref12,Ref14,Ref22,Ref24}, and typically focused on a single-user scenario~\cite{Ref12,Ref14}. In~\cite{Ref6}, an iterative heuristic algorithm was developed to maximize the EE of renewable powered hybrid arrays, subject to a data rate requirement, where antennas are selected and transmit powers are allocated in an alternating manner until convergence. In~\cite{Ref23}, the phase shifts of hybrid arrays were optimized to reduce the power consumption and improve the SE. Only a few works have taken multiuser into account, under the assumption that perfect CSI is available at the transmitter~\cite{Ref22,Ref24}. This is due to the fact that the estimation of CSI is challenging in a hybrid array, especially a localized hybrid array. The estimation accuracy can either be poor due to inherent phase ambiguity of localized hybrid arrays~\cite{Ref15,Ref16,Ref47}, or necessitate long training pilots, severe estimation delay, and accurate \mbox{a-priori} knowledge on the number of multipath components~\cite{Ref45,Ref46}. One of the few studies of multiuser in hybrid arrays is~\cite{Ref24}, where two bit-allocation algorithms were developed to minimize the quantization distortion of hybrid arrays by exploiting flexible ADC resolutions, given perfect CSI at the transmitter. Another study is~\cite{Ref9} which investigated the EE-SE trade-off and derived the most energy-efficient number of RF chains given SE.

In different yet relevant contexts, the impact of non-ideal PAs or non-negligible circuit power on the EE has been evaluated in many other wireless communication systems. In~\cite{circuitp1,circuitp2}, a string tautening algorithm was proposed to produce the most energy-efficient schedule for delay-limited traffic, first considering negligible circuit power, and then extended to non-negligible constant circuit power and energy-harvesting communications. In~\cite{Ref19}, the EE-delay trade-off of a proportionally fair downlink cellular network was studied in the case of non-ideal PAs. In~\cite{EEmetric}, the power allocation was optimized to maximize EE in conventional single-hop frequency-selective channels with non-negligible constant circuit power. In~\cite{Ref43,Ref18}, beside the transmit powers of all participating nodes, the transmit durations were optimized jointly to maximize the EE of two-way relay systems with non-ideal PAs and non-negligible circuit power. However, to our best knowledge, the compound effect of \mbox{non-ideal} PAs and non-negligible circuit power has not been considered in hybrid arrays.
\section{System Model}\label{sec:2}
\begin{figure}[!t]
\centering
\includegraphics[scale=0.32]{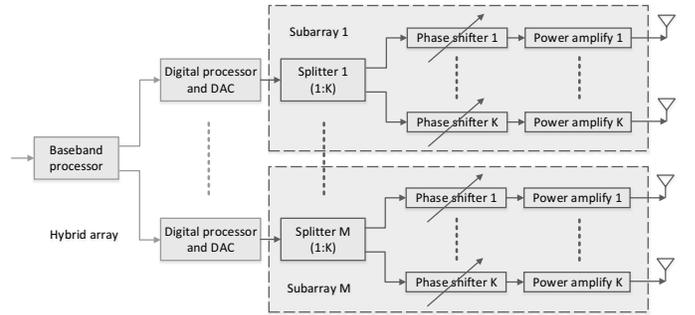}
\caption{The architecture of the hybrid array.}
\label{HybridArray}
\vspace{-4mm}
\end{figure}

We consider a slotted system with slot duration $T$, where a hybrid array is employed to transmit a stream of data for transmit duration $t$ per slot to a \mbox{single-antenna} receiver with a target average data rate, denoted by $r_{\rm dl}$. As illustrated in Fig.~\ref{HybridArray}, the hybrid array consists of $M$ analog subarrays with $K$ antennas per subarray. Each antenna has its own adjustable phase shifter, a DAC and a PA~\cite{Ref20,Ref26,Ref17}. All subarrays are connected with a baseband processor. The transmit power of an analog subarray is evenly distributed among its antennas, denoted by $p_{\rm{A}}^m$, at each antenna at the $m$-th subarray. The system bandwidth is $W$ in Hertz.

A non-ideal, non-linear power amplifier, modeled by the popular traditional power amplifier~(TPA), is connected to every antenna. The total power consumption at the PA of each antenna at the $m$-th subarray can be given by~\cite{Ref19}
\begin{equation}\label{TPA}
\Psi_{\rm{A}} (p_{\rm{A}}^m) = \frac{\sqrt{p_{\rm{A}}^m P_{\max,m}^{\rm{A}} }}{\eta_{\max,m}},\ m = 1,\cdots, M,
\end{equation}
where $P_{\max,m}^{\rm{A}}$ and $\eta_{\max,m}$ are the maximum output power and the maximum PA efficiency at each antenna at the $m$-th subarray, respectively.

Apart from the power consumed by the PAs, there is \mbox{non-negligible} power consumption in the rest of the transmitter circuit. The circuit consists of ADC/DAC, baseband processor, up-and-down converter, oscillator and so on. Part of the circuit power consumption, such as the energy consumed in the baseband processor, can be modeled to explicitly depend on the instantaneous transmit rate $R_a$, which can be written as a function of $R_a$~\cite{Ref49}, denoted by $f_p(R_a)$. As extensively considered in the literature~\cite{EEmetric,EnergyCom}, a linear function $f_p(R_a) = \epsilon_m R_a$ is adopted in this paper, where $\epsilon_m$ is the coefficient known in prior and specifies the energy consumption per bit~(in Joule per bit) at the $m$-th subarray.

The rest of the circuit power consumption can be modeled to be independent of the data rate, remain unchanged during transmission, and can be turned off after transmission, such as the energy consumed by ADC/DAC, up-and-down converter, and oscillator. It can be denoted by $P_{{\rm base},m}$ at the $m$-th subarray. As a result, the total power consumption of the $m$-th subarray can be written as
\begin{equation}\label{ptx_ori}
    P_{{\rm tx},m}= K \cdot \Psi_{\rm{A}} (p_{\rm{A}}^m) + \epsilon_{m} R_a + P_{{\rm base},m}.
\end{equation}

Further, the circuit power of the $m$-th subarray is assumed to be constant in an idle mode, denoted by $P_{{\rm idle},m}$. Without loss of generality, it is assumed that all subarrays have identical RF chains, and hence identical maximum output power $P_{\max}^{\rm A}$, maximum PA efficiency $\eta_{\max}$ and circuit power parameters, with the subscript ``$_m$'' suppressed.

Two types of beamforming techniques are considered, namely, coherent beamforming and \mbox{non-coherent} beamforming. Coherent beamforming can be conducted in the case where the CSI from all the antennas of the hybrid array to the receiver is known at the hybrid array. For example, the angles-of-departures (AoDs) are estimated by using angular search~\cite{Ref15}, extending spectral analysis~\cite{Ref45}, or conducting zero knowledge beamforming (ZKBF)~\cite{Ref44}, typically in line-of-slight~(LoS) dominant channels, prior to data transmission. The phase shifter connected to every antenna can be accordingly calibrated, so that the phases of the signals from different antennas are aligned at the receiver and constructive combination is achieved~\cite{Ref35,Ref36,Ref37}.

\mbox{Non-coherent} beamforming can be carried out in the scenario where the explicit CSI of individual antennas is unavailable to the hybrid array. Each subarray needs to independently run ZKBF~\cite{Ref44} to determine its own configuration of phase shifters until convergence. Given the local optimality of ZKBF, the convergent configuration per subarray is not necessarily optimal nor consistent among the subarrays. Space-time block coding (STBC)~\cite{Ref31,Ref32,Ref33} provides an embodiment of \mbox{non-coherent} beamforming among subarrays. The only available knowledge of the channels is the average amplitudes on a subarray basis. This information can assist the design of the most energy-efficient setting of the hybrid array under \mbox{non-coherent} beamforming, as to be described in Section~\ref{sec:3}. This scenario is of particular interest to multipath abundant environments, e.g., Rayleigh channel, where the estimation of CSI is known to be challenging and has yet to be addressed.

In terms of channel model, the algorithms proposed in this paper are general, suitable for different channel models, and not limited to any particular channel model. As extensively assumed in the literature~\cite{Ref22,Ref34}, identical and independently distributed~(i.i.d.) block Rayleigh fading channels are assumed at each antenna, which account for rich scattering environments. Let $h_m^k$ denote the channel coefficient from the $k$-th antenna of the $m$-th subarray to the single-antenna receiver. $h_m^k$ stays unchanged during a slot and changes between slots.

It is assumed that a subset of the $M$ analog subarrays, denoted by $\mathcal{M}$, transmit data $\{ s_m \}$ jointly to the receiver for $t (\leq T)$ seconds and turn into the idle mode during the rest of the slot, i.e., $(T-t)$ seconds. Therefore, the received signal at the receiver during the active time is given by
\begin{equation}\label{ReceSignal}
    y=\sum_{m \in \mathcal{M}} \sum_{k=1}^{K} \sqrt{p_{\rm{A}}^m} {h}_{m}^k s_m^k + n,
\end{equation}
where $s_m^k=\omega_m^k s_m$ is the precoded/weighted signal that the $k$-th antenna of the $m$-th subarray transmits. $\mathbb{E}\left\{ {{{\left| {{s_m^k}} \right|}^2}} \right\} = 1$. $\omega_m^k$ is the precoding coefficient for the $k$-th antenna of the $m$-th subarray. $n$ is the additive white Gaussian noise~(AWGN) at the receiver, i.e., $n \sim \mathcal{N} (0,\sigma^2)$. Let $N_0$ denote the power spectral density~(PSD) of the noise, and thus $\sigma^2 = N_0 W$.

The received power at the intended receiver can be given by
\begin{equation}\label{Useful_Power_original}
S  =
\left\{ \begin{array}{ll}
\sum\limits_{m \in \mathcal{M}} \left| \sum\limits_{k = 1}^K {\sqrt{p_{\rm{A}}^m}{{h}_{m}^k}}\right|^2, & {\text{non-coherent BF}};\\
{\left( {\sum\limits_{m \in \mathcal{M}} \sum\limits_{k = 1}^K {\sqrt{p_{\rm{A}}^m}{{\left| {h}_{m}^k \right|}}} } \right)}^2, & {\text{coherent BF}}.
\end{array} \right.
\end{equation}

By defining $p_m = K p_{\rm{A}}^m$ and
\begin{equation}\nonumber
h_m  =
\left\{ \begin{array}{ll}
\frac{1}{\sqrt{K}} \left| \sum\limits_{k=1}^{K}  h_m^k \right|, & {\text{non-coherent BF}};\\
\frac{1}{\sqrt{K}} \sum\limits_{k=1}^{K} \left| h_m^k \right|, & {\text{coherent BF}},
\end{array} \right.
\end{equation}
the received power~\eqref{Useful_Power_original} can be rewritten as
\begin{equation}\label{Useful_Power}
S  =
\left\{ \begin{array}{ll}
\sum\limits_{m \in \mathcal{M}} {{p_m}{{h_m}^2}}, & {\text{non-coherent BF}};\\
{\left( {\sum\limits_{m \in \mathcal{M}} {\sqrt {{p_m}} {h_m}} } \right)}^2, & {\text{coherent BF}}.
\end{array} \right.
\end{equation}

Therefore, the average achievable data rate during each time slot~$T$ can be given by
\begin{equation}\label{AchiRate}
    R = \frac{t}{T} \, R_a = \frac{t}{T} \, W \, \log_2 \left( 1 + \frac{S}{\sigma^2} \right),
\end{equation}
where the instantaneous data rate $R_a = W \, \log_2 \left( 1 + \frac{S}{\sigma^2} \right)$.

With the variable replacement of $\{p_m\}$, an equivalent TPA model, satisfying the total power constraint of the $m$-th subarray, can be written as
\begin{equation}\label{TPA_eq}
\Psi (p_m) = \frac{\sqrt{p_m P_{\max}} }{\eta_{\max}},\ m = 1,\cdots, M,
\end{equation}
where $P_{\max}=K P_{\max}^{\rm{A}}$ is the maximum output power of the $m$-th subarray given $K$ antennas per subarray. Therefore, the total power consumption of the $m$-th subarray can be rewritten as
\begin{equation}\label{ptx_eq}
    P_{{\rm tx},m}= \Psi(p_m) + \epsilon \, R_a + P_{{\rm base}}.
\end{equation}
\section{EE Maximization}\label{sec:3}
We aim to maximize the EE of the hybrid array under \mbox{non-ideal} PAs and \mbox{non-negligible} circuit power. The EE is defined as the ratio of the target average data rate $r_{\rm dl}$ to the average total energy consumption $E_{\rm{total}}$~\cite{EEmetric,Ref48}, as given by
\begin{equation}\label{EE_indicator}
    {\eta _E} = \frac{r_{\rm dl}}{E_{\rm{total}}/T} = \frac{{r_{\rm dl}T}}{E_{\rm{total}}}.
\end{equation}
where the second equality indicates that the EE is equivalent to the ratio between the number of bits to be transmitted within $T$ and the total energy required to transmit these bits. To this end, given the number of data bits to be sent within $T$, i.e., $r_{\rm dl}T$, maximizing EE is equivalent to minimizing $E_{\rm{total}}$.

Moreover, minimizing $E_{\rm{total}}$ facilitates maximizing the EE of the hybrid array under non-ideal PAs and non-negligible circuit power. The reason is that maximizing EE of the hybrid array may require the subarrays to transmit for part of a slot to balance the non-negligible circuit energy and PAs consumption. The optimal transmit rate may switch to null during a slot. The direct maximization of the EE, i.e., directly maximizing EE of the hybrid array, would be unsuitable, due to such change of the data rate.

With respect to $\{p_m\}$ ($m=1,\cdots,M$) and $t$, the maximization of EE can be formulated as
\begin{equation}\label{G_EE_Opt} \tag{\textbf{P1}}
\begin{aligned}
    \mathop {\min }_{\{p_m\},t} \ \ & \sum_{m = 1}^M \mathbb{I}\left( p_m \right) \Big[{P_{{\rm tx},m}} \, t + {P_{{\rm idle}}} \, \big( T - t \big) \Big]  \\
    & + \sum_{m = 1}^M \Big[ 1 - \mathbb{I}\left( p_m \right) \Big] {P_{{\rm idle}}} T, \\
    {\rm{s.t.}} \ \ & {r_{\rm dl}} \le R, \\
    & t_{\min} \le t \le T, \\
    & 0 \le \Psi \left( {{p_m}} \right) \le {P_{\max}},\ m = 1, \cdots, M,
\end{aligned}
\end{equation}
where $\mathbb{I}\left( \cdot \right)$ is an indicator function, i.e., $\mathbb{I}\left( x \right)=1$ if $x>0$; otherwise, $\mathbb{I}\left( x \right)=0$. Therefore, $\mathcal{M}=\{m:\mathbb{I}(p_i)=1,\,i=1,\cdots,M\}$, and the size of $\mathcal{M}$ is denoted by $m^*$. $t_{\min}$ is the minimum transmit duration required to meet the target of $r_{\rm dl}$, given ${P_{\max}}$. In the optimal solution for~\eqref{G_EE_Opt}, ${r_{\rm dl}} = R$ or $R_a = \frac{r_{\rm dl} T}{t}$, since $\{p_m\}$ can be continuously reduced until this equality is taken.

By defining $x_m \buildrel \Delta \over =  \frac{P_{\max}}{\eta_{\max}^2} p_m \geq 0$ and suppressing the constant term, the objective of~\eqref{G_EE_Opt} can be rewritten as
\begin{equation}\label{EE_Opt_TPA}
\sum_{m\in \mathcal{M}} \left[\sqrt{{x_m}} \, t + \big( P_{{\rm base}} - P_{{\rm idle}} \big) t\right],
\end{equation}
where, by evaluating \eqref{TPA_eq}, the feasible solution region of $x_m$ is
\begin{equation}\label{Re_PowerConstraint}
0 \le {x_m} \le P_{\max}^2.
\end{equation}

By substituting \eqref{Useful_Power} into \eqref{AchiRate} and setting ${r_{\rm dl}} = R$, the minimum data rate constraints of~\eqref{G_EE_Opt} can be rewritten as
\begin{equation}\label{Re_RateConstraint}
\left\{ {\begin{array}{*{20}{l}}
{\sum\limits_{m \in \mathcal{M}} {{x_m}\kappa _m^2 = \theta \left( t \right)} },&{{\text{non-coherent BF}};}\\
{\sum\limits_{m \in \mathcal{M}} {\sqrt {{x_m}} {\kappa _m}} = \sqrt {\theta \left( t \right)} },&{{\text{coherent BF}},}
\end{array}} \right.
\end{equation}
where $\kappa_m$, referred to as ``effective channel gain'', is given by
\begin{equation}\label{Defination_k}
{\kappa _m} = \frac{{{\eta _{\max}}}}{{\sqrt {{P_{\max}}} }} {{h_m}} >0;
\end{equation}
\begin{equation}\label{Defination_theta}
\theta \left( t \right) = \left( {{2^{\frac{{r_{\rm{dl}}}T}{tW}}} - 1} \right) {\sigma^2} >0.
\end{equation}

We assume that~\eqref{G_EE_Opt} is feasible, i.e., $t_{\min} \leq T$; in other words,
\begin{equation}\nonumber
\left\{ {\begin{array}{*{20}{l}}
{\sum\limits_{m = 1}^M {P_{\max}^2\kappa _m^2} \geq \theta \left( T \right)},&{{\text{non-coherent BF}};} \\
{\sum\limits_{m = 1}^M {{P_{\max}}{\kappa _m}} \geq \sqrt {\theta \left( T \right)}},&{{\text{coherent BF}}.}
\end{array}} \right.
\end{equation}

Unfortunately, \eqref{G_EE_Opt} is not convex due to the \mbox{non-convex} objective~\eqref{EE_Opt_TPA} under both coherent and non-coherent beamforming. This is due to the fact that the $(m^*+1) \times (m^*+1)$ Hessian matrix of~\eqref{EE_Opt_TPA}, denoted by $\mathbf{H}$, is neither positive definite nor negative definite, as given by
\begin{equation}\label{Hessian_matrix}
\mathbf{H}=\left[
  \begin{array}{ccccc}
    -\frac{1}{4} x_{\mathcal{M}(1)}^{-\frac{3}{2}} \cdot t & \cdots & 0 & \frac{1}{2} x_{\mathcal{M}(1)}^{-\frac{1}{2}} \\
    \vdots & \ddots & 0 & \vdots \\
    0 & 0 & -\frac{1}{4} x_{\mathcal{M}(m^*)}^{-\frac{3}{2}} \cdot t & \frac{1}{2} x_{\mathcal{M}(m^*)}^{-\frac{1}{2}} \\
    \frac{1}{2} x_{\mathcal{M}(1)}^{-\frac{1}{2}} & \cdots & \frac{1}{2} x_{\mathcal{M}(m^*)}^{-\frac{1}{2}} & 0 \\
  \end{array}
\right],
\end{equation}
where ${\mathcal{M}(i)}$ denotes the $i$-th subarray in $\mathcal{M}$. The feasible solution region of~\eqref{G_EE_Opt} is also non-convex due to the logarithmic data rate constraints.
\section{The Structure of Optimal Transmit Powers}\label{sec:4}
In this section, we derive the closed-form solution for the most energy-efficient transmit power of each analog subarray, given any $t \leq T$, under non-ideal PAs and non-negligible circuit power. To do this, we first arrange ${\kappa _m}$ in descending order, as given by
\begin{equation}\label{pri_order}
{\kappa_{\pi(1)}} \geq {\kappa_{\pi(2)}} \geq \cdots \geq {\kappa_{\pi(m^*)}} \geq \cdots \geq {\kappa_{\pi(M)}},
\end{equation}
where $\pi(i)$ denotes the $i$-th place in the arrangement.

We assume there are $m^*$ active analog subarrays in the hybrid array. Given identical PAs and maximum transmit powers of all subarrays, we can readily have
\begin{equation}\label{pri_fact}
x_{\pi(1)} \geq \cdots \geq x_{\pi(m^*)} \geq x_{\pi(m^*+1)}= \cdots = x_{\pi(M)} = 0.
\end{equation}
This is because, if the effective channel gains are non-consecutive, i.e., any subarray $\pi(m)$, $m<m^*$, is inactive, activating subarray $\pi(m)$ and deactivating subarray $\pi(m^*)$ would be more energy-efficient, given the higher effective channel gain of the former, i.e., $\kappa_{\pi(m)} \geq \kappa_{\pi(m^*)}$. For the same reason, if any two subarrays $\pi(i)$ and $\pi(j)$, $i,j \leq m^*$, do not meet \eqref{pri_fact}, i.e., $\kappa_{\pi(i)} \geq \kappa_{\pi(j)}$ and $x_{\pi(j)} > x_{\pi(i)}$, exchanging the values of $x_{\pi(i)}$ and $x_{\pi(j)}$ can be more energy-efficient. For details, please refer to Appendix~\ref{Apx1}

Under the TPA model and \mbox{non-negligible} circuit power, the most energy-efficient selection of subarrays is equivalent to finding $m^*$, and the subarrays with the highest $m^*$ consecutive effective channel gains, i.e., $\kappa_{\pi(i)}$, for $i=1,\cdots,m^*$, are selected to be active. Following this, {Theorem}~\ref{PowerA_TPA} provides the criterion to identify $m^*$ for both coherent and non-coherent beamforming.

\vspace{2mm}

\begin{theorem} \label{PowerA_TPA}
\textit{Given $t$, the EE of the hybrid array of interest can be maximized by turning on only the analog subarrays ${\pi(m)}$, $m \leq m^*$, with $m^*$ specified by}

\noindent \textit{for ${m^*} \geq 2$,}
\begin{equation}\label{Bset_m}
\left\{ {\begin{array}{*{20}{l}}
{\sum\limits_{m = 1}^{{m^*} \hspace{-1mm} - \hspace{-0.5mm} 1} \hspace{-1mm} {P_{\max}^2\kappa _{\pi(m)}^2} \hspace{-1mm} < \theta \left( t \right) \hspace{-1mm} \le \hspace{-2mm} \sum\limits_{m = 1}^{{m^*}} \hspace{-1mm} {P_{\max}^2\kappa _{\pi(m)}^2} }, \text{non-coherent BF};\\
{\sum\limits_{m = 1}^{{m^*} \hspace{-1mm} - \hspace{-0.5mm} 1} \hspace{-1mm} {{P_{\max}}{\kappa _{\pi(m)}}} \hspace{-1mm} < \sqrt {\theta \left( t \right)} \hspace{-1mm} \le \hspace{-2mm} \sum\limits_{m = 1}^{{m^*}} \hspace{-1mm} {{P_{\max}}{\kappa _{\pi(m)}}} }, \text{coherent BF};
\end{array}} \right.
\end{equation}
\textit{for $m^*=1$,}
\begin{equation}\label{Bset_m_1}
\left\{ {\begin{array}{*{20}{l}}
{0 < \theta \left( t \right) \le {P_{\max}^2\kappa _{\pi(1)}^2} },&{{\text{non-coherent BF}}};\\
{0 < \sqrt {\theta \left( t \right)} \le {{P_{\max}}{\kappa _{\pi(1)}}} },&{{\text{coherent BF}}}.
\end{array}} \right.
\end{equation}

\textit{The optimal transmit powers of the activated subarrays are given by}
\begin{equation}\label{Pi_TPA}
\begin{aligned}
&p_{\pi(m)}^* = {P_{\max}}\eta _{\max}^2,\ {m \in \{ 1,2,\cdots,{m^*} - 1\} },\\
&p_{{\pi(m^*)}}^* = \left\{ {\begin{array}{*{20}{l}}
{\frac{{\theta \left( t \right) - \sum\limits_{m = 1}^{{m^*} - 1} {P_{\max}^2\kappa _{\pi(m)}^2} }}{{{{\left| {{h_{{\pi(m^*)}}}} \right|}^2}}}},\text{non-coherent BF},\\
{\frac{{{{\left( {\sqrt {\theta \left( t \right)}  - \sum\limits_{m = 1}^{{m^*} - 1} {{P_{\max}}{\kappa _{\pi(m)}}} } \right)}^2}}}{{{{\left| {{h_{{\pi(m^*)}}}} \right|}^2}}}},\text{coherent BF},
\end{array}} \right.\\
&p_{\pi(m)}^* = 0,\ {m \in \{ {m^*} + 1,{m^*} + 2,\cdots,M\}},
\end{aligned}
\end{equation}
\textit{where $\sum\limits_{m = 1}^{{m^*} - 1} {P_{\max}^2\kappa _{\pi(m)}^2} = 0$ when $m^* = 1$.}
\end{theorem}

\vspace{2mm}

\noindent \begin{IEEEproof}
Please see Appendix~\ref{Apx2}.
\end{IEEEproof}
\section{Optimal Transmit Duration}\label{sec:5}
We further optimize the transmit duration $t$, based on the structure of the optimal transmit powers established in Section~\ref{sec:4}. We note that $t$ interacts with $m^*$ and the optimal transmit powers of the analog subarrays in a hybrid array.
\subsection{Feasible Region of The Transmit Duration}\label{sec:5a}
Let $t_{\min }^m$ define the minimum transmit duration that achieves the required data rate ${r_{{\rm{dl}}}}$ when there are $m (\leq M)$ active analog subarrays with the highest channel gains and transmitting the maximum transmit powers, as dictated in {Theorem}~\ref{PowerA_TPA}. Plugging $P_{\max}$ into \eqref{Re_RateConstraint} to replace $x_m, \forall m$, $t_{\min }^m$ can be resolved, as given by
\begin{equation}\label{Tmin_TPA}
t_{\min }^m = \frac{{{r_{{\rm{dl}}}}T}}{{W{{\log }_2}\left( {1 + \frac{{S_{\max }^m}}{\sigma^2}} \right)}},
\end{equation}
where
\begin{equation}\nonumber
S_{\max }^m = \left\{ {\begin{array}{*{20}{l}}
{\sum\limits_{i = 1}^m {{P_{\max}}\cdot{\eta_{\max}^2}{{\left| {{h_i}} \right|}^2}} },&{{\text{non-coherent BF}};}\\
{\left({\sum\limits_{i = 1}^m {\sqrt{P_{\max}}\cdot{\eta_{\max}}{{\left| {{h_i}} \right|}}} }\right)}^2,&{{\text{coherent BF}},}
\end{array}} \right.
\end{equation}
and
\begin{equation}\nonumber
t_{\min }^{M}<t_{\min }^{M-1}< \cdots < t_{\min }^{m} < \cdots < t_{\min }^{1}.
\end{equation}
Apparently, \eqref{G_EE_Opt} is infeasible if $t_{\min }^M>T$. We consider the case with a \mbox{non-empty} feasible solution region, i.e., $t_{\min }^M \leq T$.

We can partition the feasible solution region into the following $M$ segments:
\begin{equation}\nonumber
\left[\min \{t_{\min }^{m},T\},\min \{t_{\min }^{m-1},T\}\right),\ m \in \{1,2,\cdots,M\},
\end{equation}
where $t_{\min }^{0}=T$, and the feasible solution region $\left[\min \{t_{\min }^{m},T\},\min \{t_{\min }^{m-1},T\}\right) = \emptyset$ if $t_{\min }^{m} \geq T$.

For any non-empty feasible solution region $t \in [t_{\min }^{m},\min \{t_{\min }^{m-1},T\}) \subseteq [t_{\min }^{m},t_{\min }^{m-1}),\ m \in \{2,3,\cdots,M\}$, the following relationship withstands:
\begin{equation}\label{Re_Region_Tmin_TPA}
{S_{\max }^{m-1}}< \theta \left( t \right) \leq {S_{\max }^m},
\end{equation}
which, satisfying \eqref{Bset_m}, indicates that subarrays $\pi(i)$, $i=1,\cdots,m$, are turned on.

If $t_{\min }^{1} < T$, for $t \in [t_{\min }^{1},\min \{t_{\min }^{0},T\}) = [t_{\min }^{1},T)$, the following relationship withstands:
\begin{equation}\label{Re_Region_Tmin_Spec}
0 < \theta \left( T \right) < \theta \left( t \right) \leq {S_{\max }^1},
\end{equation}
which, also satisfying \eqref{Bset_m}, indicates that subarry $\pi(1)$ alone is turned on.
\subsection{Optimization Reformulation and Solution}\label{sec:5b}
For each of the above segments of the feasible solution region, i.e., $t \in [t_{\min }^{m},\min \{t_{\min }^{m-1},T\})$ with $t_{\min }^{m} < T$, if ${m = 1}$, the most energy-efficient transmit power of subarray $\pi(1)$ is given by~\eqref{Pi_TPA} and the rest of the subarrays are turned off; otherwise, if $m \geq 2$, the most energy-efficient transmit powers of subarrays $\pi(i)$, $i=1,\cdots,m-1$, are ${P_{\max ,{\pi(i)}}}\eta _{\max ,{\pi(i)}}^2$, the transmit power of subarray $\pi(m)$ can also be given by~\eqref{Pi_TPA}, and the rest of the subarrays are turned off, as dictated in {Theorem}~\ref{PowerA_TPA}. With $t$ being the only variable to be determined, optimization problem~\eqref{G_EE_Opt} can be reformulated over the segment, as given by
\begin{equation}\label{Re_Opt_TPA}\tag{\textbf{P2}}
\begin{aligned}
{\min\limits_t} \ E_{\rm total}^{m}(t) &= \frac{{\sqrt {{P_{\max}}} }}{{{\eta _{\max}}}}\sqrt {p_m^*} t + \sum\limits_{i = 1}^{m - 1} {{P_{\max}}} t \\
& + \sum\limits_{i = 1}^m ( {P_{{\rm{base}}}} - {P_{{\rm{idle}}}}) t, \\
{\rm{s.t.}} \ \ & t_{\min }^{m} \leq t < \max \{t_{\min }^{m-1},T\},
\end{aligned}
\end{equation}
where ${p_m^*}$ is referred to~\eqref{Pi_TPA}, and $\sum\limits_{i = 1}^{m - 1} {{P_{\max}}} t = 0$ when ${m = 1}$.

\vspace{2mm}

\begin{theorem} \label{Convex_Convave}
\textit{In the case of \mbox{non-coherent} beamforming, \eqref{Re_Opt_TPA} is concave in $[t_{\min }^{m},\min \{t_{\min }^{m-1},T\}),m \in \{2,\dots,M\}$, and convex in $[t_{\min }^{1},T]$ if $\frac{r_{\rm{dl}}}{W} \geq 1$. In the case of coherent beamforming, \eqref{Re_Opt_TPA} is convex if $\frac{r_{\rm{dl}}}{W} \geq 1$.}
\end{theorem}

\vspace{2mm}

\noindent \begin{IEEEproof}
In the case of \mbox{non-coherent} beamforming, according to
\begin{equation}\label{SencondOrder_Non}
\frac{{{\partial ^2}E_{{\rm{total,1}}}^m(t)}}{{\partial {t^2}}} = \frac{{{2^{\frac{{r_{{\rm{dl}}}}T}{tW} - 2}} {\sigma^2} \upsilon {{\left( {\frac{{{r_{{\rm{dl}}}}T}}{W}} \right)}^2}{{\left( {\ln 2} \right)}^2}}}{{{{\left[ {\upsilon + \sum\limits_{i = 1}^{m - 1} {P_{\max}^2\kappa _i^2} } \right]}^{3/2}} {{h_m}}{t^3}}},
\end{equation}
the sign of the second-order derivative of $E_{\rm total}^{m}$, i.e., $\frac{{{\partial ^2}E_{{\rm{total,1}}}^m(t)}}{{\partial {t^2}}}$, is determined by
\begin{equation}\nonumber
\upsilon = {\left( {{2^{\frac{{r_{{\rm{dl}}}}T}{tW}}} - 2} \right) {\sigma^2} - 2\sum\limits_{i = 1}^{m - 1} {P_{\max}^2\kappa _i^2}}.
\end{equation}

From \eqref{Pi_TPA}, we have
\begin{equation}\nonumber
\left( {{2^{\frac{{r_{{\rm{dl}}}}T}{tW}}} - 1} \right) \sigma^2 - \sum\limits_{i = 1}^{m - 1} {P_{\max}^2\kappa _i^2} = {x_m}\kappa _m^2 > 0,
\end{equation}
based on which $\upsilon$ can be rewritten as
\begin{equation}\nonumber
\upsilon = {x_m}\kappa _m^2 - \sum\limits_{i = 1}^{m - 1} {P_{\max}^2\kappa _i^2} - {\sigma^2}.
\end{equation}

For $t \in [t_{\min }^{m},\min \{t_{\min }^{m-1},T\})$, $2 \leq m \leq M$ and $t_{\min }^{m} < T$, according to~\eqref{pri_order} and~\eqref{pri_fact}, we know
\begin{equation}\nonumber
0 < \kappa _m \leq \kappa _i, i \in \{1,2,\dots,m-1\},
\end{equation}
\begin{equation}\nonumber
0 < x_m \leq x_i \leq P_{\max}^2, i \in \{1,2,\dots,m-1\},
\end{equation}
from which we can have
\begin{equation}\nonumber
{x_m}\kappa _m^2 - \sum\limits_{i = 1}^{m - 1} {P_{\max}^2\kappa _i^2}<0.
\end{equation}
It is concluded that $\upsilon<0$, and in turn $\frac{{{\partial ^2}E_{{\rm{total,1}}}^m(t)}}{{\partial {t^2}}}<0$. With the linear constraint, \eqref{Re_Opt_TPA} is concave in $[t_{\min }^{m},\min \{t_{\min }^{m-1},T\})$, $m \in \{2,\dots,M\}$.

For $t \in [t_{\min }^{1},T] \neq \emptyset$, $t_{\min }^{1} < T$, according to {Theorem}~\ref{PowerA_TPA}, $\sum\limits_{i = 1}^{m - 1} {P_{\max ,i}^2\kappa _i^2}=0$ for $m=1$, based on which $\upsilon$ can be rewritten as
\begin{equation}\nonumber
\upsilon = {\left( {{2^{\frac{{r_{{\rm{dl}}}}T}{tW}}} - 2} \right) {\sigma^2}}.
\end{equation}
Clearly, $\upsilon$ is positive if $\frac{r_{\rm{dl}}}{W} \geq 1$. Therefore, $\frac{{{\partial ^2}E_{{\rm{total,1}}}^m(t)}}{{\partial {t^2}}}>0$ and \eqref{Re_Opt_TPA} is convex in $[t_{\min }^{1},T]$ if $\frac{r_{\rm{dl}}}{W} \geq 1$.

In the case of coherent beamforming, according to
\begin{equation}\label{SencondOrder_Coh}
\frac{{{\partial ^2}E_{{\rm{total,2}}}^m(t)}}{{\partial {t^2}}} = \frac{{{2^{\frac{{r_{{\rm{dl}}}}T}{tW} - 2}}\left( {{2^{ \frac{{r_{{\rm{dl}}}}T}{tW}}} - 2} \right){\sigma^4}{{\left( {\frac{{{r_{{\rm{dl}}}}T}}{W}} \right)}^2}{{\left( {\ln 2} \right)}^2}}}{{{{\left[ {\left( {{2^{\frac{{r_{{\rm{dl}}}}T}{tW}}} - 1} \right) {\sigma^2}} \right]}^{3/2}} {{h_m}} {t^3}}},
\end{equation}
like the case of $t \in [t_{\min }^{1},T]$ under \mbox{non-coherent} beamforming, it can be readily concluded that $\frac{{{\partial ^2}E_{{\rm{total,2}}}^m(t)}}{{\partial {t^2}}}>0$ if $\frac{r_{\rm{dl}}}{W} \geq 1$ is satisfied. As a result, \eqref{Re_Opt_TPA} is proved to be convex in the case of coherent beamforming if $\frac{r_{\rm{dl}}}{W} \geq 1$.
\end{IEEEproof}
\subsubsection{Optimization under Non-coherent Beamforming}\label{sec:5c}
By {Theorem}~\ref{Convex_Convave}, the optimal solution for~\eqref{G_EE_Opt} can be achieved by comparing the solutions for~\eqref{Re_Opt_TPA} in different segments of the feasible solution region of $t$. The optimal solution for~\eqref{Re_Opt_TPA} in each of the segments can be readily solved, as follows.

In the case that $t_{\min }^{1}>T$, i.e., $[t_{\min }^{1},T]=\emptyset$, the global optimal solution for~\eqref{G_EE_Opt} can be given by
\begin{equation}\label{SelectCriterion1_TPA}
t^*_{\rm tpa} \hspace{-1mm} = \hspace{-1mm} \arg\min_t \, \{E_{\rm total}^{m} \hspace{-1mm} \left(t_{\min }^{m}\right) \hspace{-1mm}, E_{\rm total}^{\bar{m}} \hspace{-1mm} \left(T\right)\},\ m = \bar{m},\cdots,M,
\end{equation}
since~\eqref{Re_Opt_TPA} has its optimum on the boundary of the feasible solution region $\left[\min \{t_{\min }^{m},T\},\min \{t_{\min }^{m-1},T\}\right)$. Here, $\bar{m}$ depends on the first non-empty feasible solution region: $\left[\min \{t_{\min }^{\bar{m}},T\},\min \{t_{\min }^{\bar{m}-1},T\}\right) \neq \emptyset$ and $\left[\min \{t_{\min }^{m},T\},\min \{t_{\min }^{m-1},T\}\right) = \emptyset$ if $m = 1,2,\cdots,\bar{m}-1$.

In the case that $t_{\min }^{1} \leq T$, i.e., $[t_{\min }^{1},T] \neq \emptyset$, \eqref{Re_Opt_TPA} is convex in $[t_{\min }^{1},T]$, and the optimal solution for~\eqref{G_EE_Opt} is either taken on the boundary of each segment of the feasible solution region, or at the fixed point of~\eqref{Re_Opt_TPA} within $[t_{\min }^{1},T]$. The fixed point, denoted by $E_{\rm total}^{1}$, can be obtained by using standard convex methods, e.g., the linear search method (as adopted in this paper). By comparing these local optimal solutions, the global optimal solution for~\eqref{G_EE_Opt} can be given by
\begin{equation}\label{SelectCriterion2_TPA}
t^*_{\rm tpa} = \arg\min_t \, \{E_{\rm total}^{m}\left(t_{\min }^{m}\right),E_{\rm total}^{1}\},\ m=2,\cdots,M.
\end{equation}
\subsubsection{Optimization under Coherent Beamforming}\label{sec:5d}
In the case of coherent beamforming, \eqref{Re_Opt_TPA} is convex as long as $\frac{r_{\rm{dl}}}{W} \geq 1$, as dictated in~{Theorem}~\ref{Convex_Convave}. We can find $\bar{m}$ to satisfy $t_{\min }^{M}<t_{\min }^{M-1}< \dots <t_{\min }^{\bar{m}} < T$, and $T \leq t_{\min }^{\bar{m}-1}$ if $\bar{m}>1$. The optimal solution is within $[t_{\min }^{m},t_{\min }^{m-1}),m \in \{\bar{m}+1,\dots,M\}$ or $[t_{\min }^{\bar{m}},T]$. For $[t_{\min }^{m},t_{\min }^{m-1})$, \eqref{Re_Opt_TPA} is convex and can be solved by the linear search method. The optimal solution in the segment, denoted by $E_{\rm total}^{m}, m \in \{\bar{m}+1,\dots,M\}$, can be obtained. For $[t_{\min }^{\bar{m}},T]$, the optimal solution, denoted by $E_{\rm tpa}^{\bar{m}}$, can be obtained in the same way. Comparing these solutions, the global optimal solution for~\eqref{G_EE_Opt} can be achieved, as given by
\begin{equation}\label{SelectCriterion3_TPA}
t^*_{\rm tpa} = \arg\min_t \, \{E_{\rm total}^{m}\},\ m \in \{\bar{m},\dots,M\}.
\end{equation}

The optimal number of active subarrays, denoted by $m^*_{\rm tpa}$, and the optimal transmit powers $p_m^*$, can be achieved along with $t^*_{\rm tpa}$, by exploiting {Theorem}~\ref{PowerA_TPA}. Following the above discussions, Algorithms \ref{alg1} and \ref{alg2} are summarized to solve \eqref{G_EE_Opt} under non-coherent and coherent beamforming, respectively. As shown in the algorithms, linear search is carried out across the $K$ antennas of a subarray, for each subarray $m \leq M$. Given a total of $M$ subarrays at the hybrid array, the worst-case complexities of the proposed algorithms are $\mathcal{O}(M\times K)$.

\begin{algorithm}[t]
\caption{Non-coherent beamforming}
\label{alg1}
\begin{algorithmic}[1]
\STATE Given $|\sum\limits_{k=1}^{K}  h_m^k |$, $\forall m$, calculate all $h_m$;
\STATE Calculate ${\kappa _m},m \in \{1,2,\cdots,M\}$ using~\eqref{Defination_k}, and arrange the $M$ subarrays in the descending order of ${\kappa _m}$;
\STATE Calculate $t_{\min }^{m}, \forall m$ with \eqref{Tmin_TPA}, and obtain all the feasible regions;
\IF{$t_{\min }^{M}>T$}
    \STATE The problem is infeasible and the algorithm terminates;
\ENDIF
\STATE Find the first non-empty feasible region and record $\bar{m}$;
\IF{$\bar{m}>1$}
        \FOR{$m = M : \bar{m}$}
            \STATE Compute $E_{\rm total}^{m}\left(t_{\min }^{m}\right)$ where subarrays $\pi(1), \cdots$, and $\pi(m)$ transmit with $P_{\max}$;
        \ENDFOR
        \STATE Compute $E_{\rm total}^{\bar{m}}\left(T\right)$ where the transmit powers are obtained from \eqref{Pi_TPA};
        \STATE Select optimal $t^*_{\rm tpa}$ using~\eqref{SelectCriterion1_TPA} and record $m^*$ and $p_{m}^*$;
\ELSIF{$\bar{m}=1$}
        \FOR{$m = M : 1$}
            \STATE Compute $E_{\rm total}^{m}\left(t_{\min }^{m}\right)$ where subarrays $\pi(1), \cdots$, and $\pi(m)$ transmit with $P_{\max}$;
        \ENDFOR
        \STATE Optimize $t$ in~\eqref{Re_Opt_TPA} for $t \in [t_{\min }^{1},T]$ by liner search and record the optimal value $E_{\rm total}^{1}$;
        \STATE Select optimal $t^*_{\rm tpa}$ using~\eqref{SelectCriterion2_TPA} and record $m^*$ and $p_{m}^*, m \in \{1,\dots,m^*\}$;
\ENDIF
\end{algorithmic}
\end{algorithm}

\begin{algorithm}[t]
\caption{Coherent beamforming}
\label{alg2}
\begin{algorithmic}[1]
\STATE Estimate channel gain $h_m^k$ for $\forall m,k$ between the hybrid array and the receiver, and calculate all $h_m$;
\STATE Calculate ${\kappa _m},m \in \{1,2,\cdots,M\}$ using~\eqref{Defination_k}, and arrange the $M$ subarrays in the descending order of ${\kappa _m}$;
\STATE Calculate $t_{\min }^{m}, \forall m$ with \eqref{Tmin_TPA}, and obtain all the feasible regions;
\IF{$t_{\min }^{M}>T$}
    \STATE The problem is infeasible and the algorithm terminates;
\ENDIF
\STATE Find the first non-empty feasible region and record $\bar{m}$;
\FOR{$m = M : (\bar{m}+1)$}
    \STATE Run linear search to optimize $t$ in \eqref{Re_Opt_TPA}, where the transmit powers are obtained from \eqref{Pi_TPA}, for $t \in [t_{\min }^{m},t_{\min }^{m-1}]$, and record the optimal value $E_{\rm total}^{m}$;
\ENDFOR
\STATE Run linear search to optimize $t$ in \eqref{Re_Opt_TPA} for $t \in [t_{\min }^{\bar{m}},T]$ and record the optimal value $E_{\rm total}^{\bar{m}}$, where the transmit powers are obtained from \eqref{Pi_TPA};
\STATE Select optimal $t^*_{\rm tpa}$ using~\eqref{SelectCriterion3_TPA} and record $m^*$ and $p_{m}^*, m \in \{1,\dots,m^*\}$;
\end{algorithmic}
\end{algorithm}

\subsection{Discussion and Extension}\label{sec:5e}
Our analysis can be extended in a more general scenario, where $f_p(\cdot)$ is unnecessarily linear, and the overall circuit power at each subarray is given by
\begin{equation}\nonumber
    P_{{\rm cir}}= P_{{\rm base}} + f_p(R_a) = P_{{\rm base}} + f_p(\frac{r_{\rm dl} T}{t}),
\end{equation}
which is convex with respect to $t$ due to its positive second-derivative, as given by
\begin{equation}\nonumber
    \frac{\partial^2 f_p(R_a)}{\partial t^2} = \frac{\partial^2 f_p(R_a)}{\partial R_a^2} (\frac{\partial R_a}{\partial t})^2 + \frac{\partial f_p(R_a)}{\partial R_a} \frac{\partial^2 R_a}{\partial t^2} > 0,
\end{equation}
where $\frac{\partial f_p(R_a)}{\partial R_a} > 0$, since the higher $R_a$ is, the more energy the circuit consumes; and $\frac{\partial^2 f_p(R_a)}{\partial R_a^2} > 0$ as it is reasonable for this part of circuit power consumption to grow increasingly faster with, if not linearly to, $R_a$. The optimization of the transmit power would be unaffected. The optimization of the transmit duration would change to
\begin{equation}\nonumber
\begin{aligned}
{\min\limits_t} \ E_{\rm total}^{m}(t) &= \frac{{\sqrt {{P_{\max}}} }}{{{\eta _{\max}}}}\sqrt {p_m^*} t + \sum\limits_{i = 1}^{m - 1} {{P_{\max}}} t \\
&+ \sum\limits_{i = 1}^m [ {f_p(\frac{ r_{\rm dl} T}{t}) + P_{{\rm{base}}}} - {P_{{\rm{idle}}}}] t.
\end{aligned}
\end{equation}

Under coherent beamforming, this can also be solved by using standard convex techniques, due to its convexity, as evident from
\begin{equation}\nonumber
\begin{aligned}
\frac{{{\partial ^2}E_{{\rm{total,2}}}^m(t)}}{{\partial {t^2}}} &= \frac{{{2^{\frac{{r_{{\rm{dl}}}}T}{tW} - 2}}\left( {{2^{ \frac{{r_{{\rm{dl}}}}T}{tW}}} - 2} \right){\sigma^4}{{\left( {\frac{{{r_{{\rm{dl}}}}T}}{W}} \right)}^2}{{\left( {\ln 2} \right)}^2}}}{{{{\left[ {\left( {{2^{\frac{{r_{{\rm{dl}}}}T}{tW}}} - 1} \right) {\sigma^2}} \right]}^{3/2}} {{h_m}} {t^3}}} \\
&+ \sum\limits_{i = 1}^m \frac{\partial^2 f_p(R_a)}{\partial R_a^2} \frac{r_{\rm dl}^2 T^2}{t^3} > 0,
\end{aligned}
\end{equation}
where the first term on the right-hand side~(RHS) of the equality can be proved to be positive in the same way as in~\eqref{SencondOrder_Coh}, and the second term is positive, as discussed earlier. The optimal solution can be achieved in the same way as described in Algorithm~\ref{alg2}.

Under \mbox{non-coherent} beamforming, we have
\begin{equation}\nonumber
\begin{aligned}
\frac{{{\partial ^2}E_{{\rm{total,1}}}^m(t)}}{{\partial {t^2}}} &= \frac{{{2^{\frac{{r_{{\rm{dl}}}}T}{tW} - 2}} {\sigma^2} \upsilon {{\left( {\frac{{{r_{{\rm{dl}}}}T}}{W}} \right)}^2}{{\left( {\ln 2} \right)}^2}}}{{{{\left[ {\upsilon + \sum\limits_{i = 1}^{m - 1} {P_{\max}^2\kappa _i^2} } \right]}^{3/2}} {{h_m}}{t^3}}} \\
&+ \sum\limits_{i = 1}^m \frac{\partial^2 f_p(R_a)}{\partial R_a^2} \frac{r_{\rm dl}^2 T^2}{t^3}.
\end{aligned}
\end{equation}
For $t \in [t_{\min }^{1},T] \neq \emptyset$, $t_{\min }^{1} < T$, we have $\frac{{{\partial ^2}E_{{\rm{total,1}}}^m(t)}}{{\partial {t^2}}} > 0$ because the first term on the RHS of the equality can be proved to be positive in the same way as in~\eqref{SencondOrder_Non}, and the second term is positive.

For $t \in [t_{\min }^{m},\min \{t_{\min }^{m-1},T\})$, $2 \leq m \leq M$ and $t_{\min }^{m} < T$, the first term on the RHS can be proved to be negative in the same way as in~\eqref{SencondOrder_Non}, while the second term is positive. Nevertheless, given the parameters $r_{\rm dl}$, $T$, $W$, $\sigma^2$, $R_a$, and $f_p(\cdot)$, we can numerically evaluate the sign of $\frac{{{\partial ^2}E_{{\rm{total,1}}}^m(t)}}{{\partial {t^2}}}$. In the case that the sign remains non-negative, $E_{\rm{total},1}^m(t)$ is convex over the segment $t\in[t_{\min}^m,\min\{t_{\min}^{m-1},T\})$. In the case that the sign remains non-positive, $E_{\rm{total},1}^m(t)$ is concave over the segment. In the case that $\frac{{{\partial ^2}E_{{\rm{total,1}}}^m(t)}}{{\partial {t^2}}}=0$ can have one or multiple roots within the segment, the segment can be further partitioned by the roots and each of the subdivided segments still yields self-contained convexity or concavity. By taking Algorithm~\ref{alg1}, the global optimal solution for $E_{\rm{total},1}^m(t)$  can be resolved efficiently by evaluating the boundaries of all resultant segments and the fixed points of those yielding convexity.

The single-user scenario that we consider is of practical value and has a range of important applications, such as satellite communications, where circuitry and PAs are \mbox{non-ideal}, and EE is critical. Particularly, the approach developed under coherent beamforming provides strong beamforming gain and high EE, provided precise CSI can be estimated by using techniques such as those proposed in~\cite{Ref15,Ref45,Ref44} in LoS dominant environments, e.g., satellite communications. The approach developed under \mbox{non-coherent} beamforming corresponds to the more realistic case where CSI may not be accurate at the transmitter, e.g., in the presence of a large number of scatters. Equally important is a multiuser scenario which can involve different beamforming techniques, and therefore can be \mbox{non-trivial} in the presence of inaccurate CSI, \mbox{non-negligible} circuit power, and \mbox{non-ideal} PAs. Significant effort would be required. The multiuser scenario will be the focus of our future work.
\section{Simulation and Numerical Results}\label{sec:7}
Simulations are carried out to validate our EE maximization of hybrid arrays with non-ideal PAs and non-negligible circuit power. Apart from the proposed algorithms, i.e., Algorithms \ref{alg1} and \ref{alg2}, we also simulate the following state-of-the-art approaches for comparison purpose.
\begin{itemize}
  \item Fixed scheme: All subarrays are active with uniformly allocated transmit powers, and the hybrid array transmits all the time.
  \item Optimized transmit duration: All subarrays are active with uniformly allocated transmit powers, and the transmit duration is optimized, as done in~\cite{Ref43,Ref18}
  \item Water-filling: All subarrays transmit over the optimized transmit duration, and the transmit powers of the subarrays are optimized by the water-filling algorithm.
\end{itemize}

For fair comparison, we ensure that all these schemes have the same required data rate $r_{\rm dl}$ over $T$. Other simulation parameters are listed in Table~\ref{tab:1}.

\begin{table}[b]
\vspace{-3mm}
\centering
\caption{Simulation Parameters}
\vspace{-2mm}
\begin{tabular}{l|l}
\hline
\textbf{Parameters} & \textbf{Values} \\
\hline
System bandwidth ($W$) & 10\,MHz \\
Time slot duration ($T$) & 10\,ms \\
Noise power spectral density ($N_0$) & --\,174\,dBm/Hz \\
Small-scale path loss & Rayleigh fading \\
Transmission distance ($d$) & 200\,m \\
Omnidirectional path loss ($PL$) & $61.4 \hspace{-0.5mm} + \hspace{-0.5mm} 20\log_{10}(d) \hspace{-0.5mm} + \hspace{-0.5mm} \xi$ \,dB \\
Lognormal shadowing of channel ($\xi$) & $\xi \sim \mathcal{N}(0,5.8^2)$ \,dB \\
Idle power consumption ($P_{{\rm idle}}$) & 30\,mW \\
Static circuit power ($P_{{\rm base}}$) & 50\,mW \\
Dynamic circuit coefficient ($\epsilon$) & 5\,mW/Mbps \\
Maximum output power ($P_{\max}$) & 46\,dBm \\
Maximum PA efficiency ($\eta_{\max}$) & 0.35\\
\hline
\end{tabular}
\label{tab:1}
\end{table}

\begin{figure}[!t]
\centering
\subfloat[Coherent beamforming]{\hspace{-4mm} \label{fig2}
\includegraphics[scale=0.28]{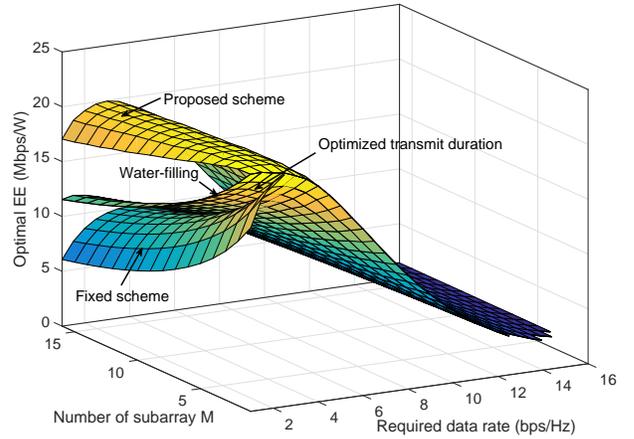}}
\vspace{-4mm}
\hfil
\centering
\subfloat[Non-coherent beamforming]{\hspace{-4mm} \label{fig3}
\includegraphics[scale=0.28]{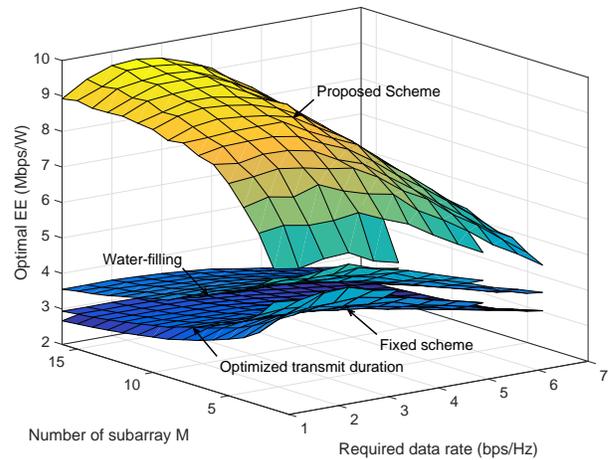}}
\caption{The optimal EE of hybrid array with $16$ antennas per subarrary considering TPA and non-negligible circuit power ($K=16$).}
\label{fig23}
\vspace{-4mm}
\end{figure}
Fig.~\ref{fig23} plots the optimal EE of hybrid arrays with $M$ analog aubarrays and $16$ antennas per subarrary, where $M$ ranges from 2 to 16. Generally, coherent beamforming can achieve higher EE than its \mbox{non-coherent} counterpart by exploiting the availability of explicit CSI. It is also clear that the proposed approach can outperform the benchmarks in both coherent and \mbox{non-coherent} beamforming. Nevertheless, the gains of the proposed algorithms decline as $r_{\rm dl}$ increases. This is due to the fact that all subarrays need to be activated and transmit with the maximum transmit power $P_{\max}$ over $T$ to support the high data rate requirement. Increasing the number of subarrays can slow down this decline, since more transmit powers are involved and can be optimized.

In the case of coherent beamforming, Fig.~\ref{fig2} shows that the EE, maximized by Algorithm~\ref{alg2}, decreases with $r_{\rm dl}$, when $M$ is small. This is due to the fact that the transmit power increases exponentially to achieve the linear growth of the data rate, hence compromising the EE. However, the EE increases first and then decreases, when $M$ is large. This is the case that the circuit power dominates over the transmit power. Particularly, in a low data rate region, the EE decreases as $M$ increases, because only a small set of subarrays need to be activated for transmission. An increased number of subarrays would lead to an increased number of idle subarrays consuming the circuit power. It is observed that the curves of the optimized transmit duration scheme and the water-filling scheme overlap under coherent beamforming. The reason is that both schemes exploit precise CSI, correct phase offsets between antennas, and achieve constructive combination of transmitted signals at the intended receiver.

In the case of \mbox{non-coherent} beamforming, Fig.~\ref{fig3} shows that the EEs of all schemes increase first and then decrease with the growth of $r_{\rm dl}$, for the same reason underlying coherent beamforming. Unlike coherent beamforming though, the EE can be improved by increasing the number of subarrays under \mbox{non-coherent} beamforming. The reason is that an increasing number of subarrays can lead to the growth of diversity in regards of the channels of all antennas and subarrays. This can lead to the increasing effectiveness of subarray selection to save energy and improve EE. Despite the growing number of subarrays raises the circuit energy consumption, the increasingly saved transmission energy resulting from the growth of diversity, can outgrow and compensate for the increasing circuit energy consumption.

\begin{figure}[!t]
\centering
\includegraphics[scale=0.28]{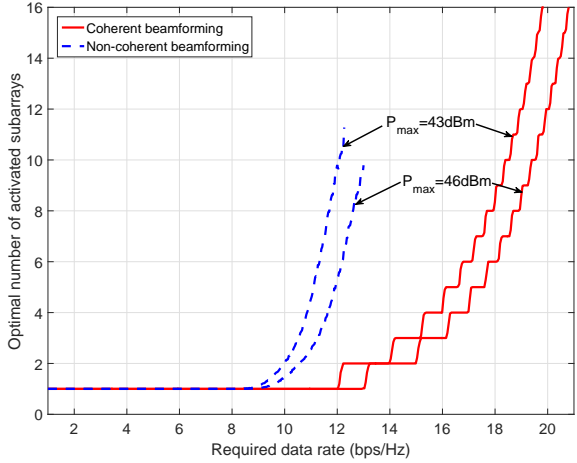}
\vspace{-8mm}
\caption{The optimal number of active subarrays for the proposed algorithms considering both TPA and non-negligible circuit power ($M = 16$ and $K = 16$).}
\label{fig4}
\vspace{-4mm}
\end{figure}
Fig.~\ref{fig4} plots the average number of active subarrays optimized by the proposed algorithms in a hybrid array with $16$ subarrays and $16$ antennas per subarray. As expected, coherent beamforming can support higher data rate requirement than \mbox{non-coherent} beamforming as the result of the availability of explicit CSI at the transmitter. We see that, given $r_{\rm dl}$, the least number of subarrays are turned on to reduce PA and non-negligible circuit power consumptions. As $r_{\rm dl}$ increases, the subarrays are increasingly activated. \mbox{Non-coherent} beamforming activates more subarrays than coherent beamforming. In other words, the lack of explicit CSI needs to be compensated for by a large number of subarrays.

It is interesting to note that the average number of active subarrays grows continuously in the case of \mbox{non-coherent} beamforming, but in a discontinuous fashion in the case of coherent beamforming. The reason is that the increasing, randomness bearing diversity drives the growth of data rate under \mbox{non-coherent} beamforming. In contrast, the growing total transmit power of the increasing number of subarrays drives the growth under coherent beamforming.

\begin{figure}[!t]
\centering
\includegraphics[scale=0.28]{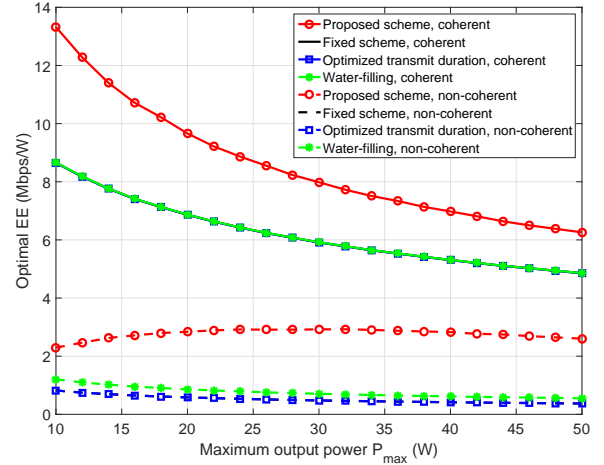}
\vspace{-8mm}
\caption{The effect of maximum output power $P_{\max}$ on optimal EE of hybrid array considering both TPA and non-negligible circuit power ($M = 16$, $K = 16$ and $r_{\rm dl}=100$ Mbps).}
\label{fig9}
\vspace{-4mm}
\end{figure}
Fig.~\ref{fig9} shows the impact of $P_{\max}$ on the optimal EE, where $r_{\rm dl} = 100$ Mbps. In the case of coherent beamforming, the EE of the proposed algorithm decreases monotonically with the growth of $P_{\max}$, due to the increasing PA consumption. In the case of \mbox{non-coherent} beamforming, the EE of the proposed algorithm first increases and then decreases for the following reason. When $P_{\max} < 30$ Watts is small, increasing $P_{\max}$ helps reduce the number of active subarrays. The energy that can be correspondingly saved is higher than the extra energy consumed at the \mbox{non-ideal} PAs. When $P_{\max} > 30$ Watts is large, more energy is consumed at the PAs than saved from reducing the number of active subarrays.

\begin{figure}[!t]
\centering
\includegraphics[scale=0.28]{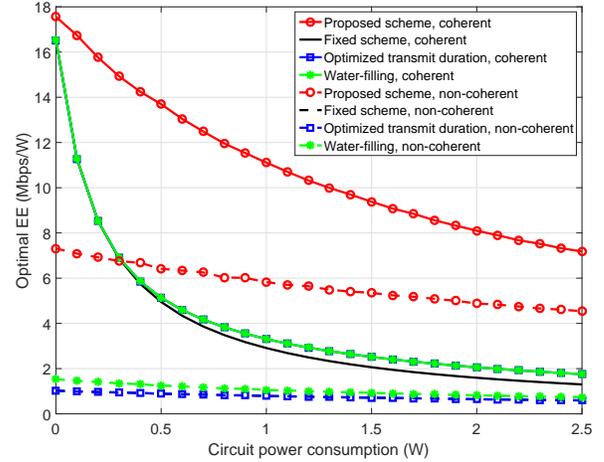}
\vspace{-8mm}
\caption{The effect of circuit power consumption on optimal EE of hybrid array with TPA ($M = 16$, $K = 16$, and $r_{\rm dl}=60$ Mbps).}
\label{fig10}
\vspace{-4mm}
\end{figure}
Fig.~\ref{fig10} shows the impact of \mbox{non-negligible} circuit power on the optimal EE, where $r_{\rm dl}=60$ Mbps. We see that the EE of the proposed scheme decreases as the circuit power consumption increases under both coherent and \mbox{non-coherent} beamforming. However, the EE of \mbox{non-coherent} beamforming decreases much more slowly, since much larger transmit powers are required due to poor equivalent channels. Moreover, the EE of the proposed algorithm decreases more slowly than those of the benchmarks, since a less number of subarrays are activated in the proposed algorithm and the total circuit power consumptions is lower under the proposed algorithm.

It is pointed out that, in Fig.~\ref{fig9}, the curves of the fixed and optimized transmit duration schemes are overlapped for both coherent and \mbox{non-coherent} beamforming. This is because $r_{\rm dl} = 100$ Mbps is high and the optimized transmit duration scheme has to transmit for the entire slot $T$ to meet $r_{\rm dl}$, as the fixed scheme does. The same reason applies to \mbox{non-coherent} beamforming in Fig.~\ref{fig10}. Nevertheless, the EE gain of the optimized transmit duration scheme over its fixed counterpart emerges and becomes conspicuous for $r_{\rm dl} = 60$ Mbps, especially under coherent beamforming. In other words, as $r_{\rm dl}$ decreases, the optimal transmit duration becomes increasingly likely to be less than $T$, and reduces energy consumption, as compared to continuous transmissions throughout $T$. Moreover, in Figs.~\ref{fig9} and~\ref{fig10}, the curves of the optimized transmit duration scheme and the classical water-filling scheme are overlapped under coherent beamforming. The reason is that both the schemes exploit precise CSI, correct phase offsets between antennas, and achieve constructive combination of transmitted signals at the intended receiver, as discussed in~Fig.~\ref{fig2}.

\begin{figure}[!t]
\centering
\includegraphics[scale=0.28]{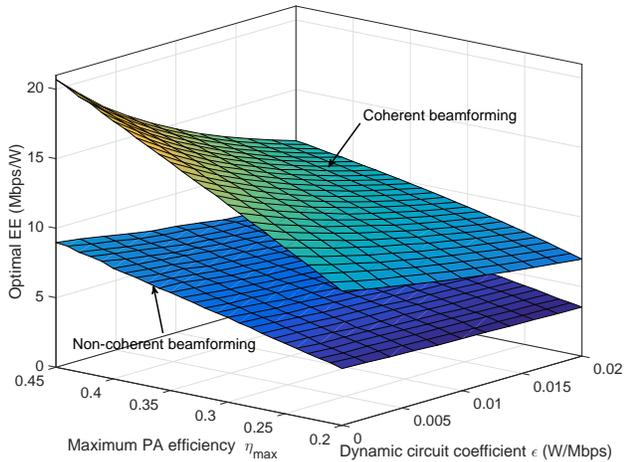}
\vspace{-8mm}
\caption{The optimal EE of hybrid arrays with TPA and non-negligible circuit power for different $\eta_{\max}$ and $\epsilon$ ($M = 16$, $K = 16$, and $r_{\rm dl}=60$ Mbps).}
\label{fig1}
\vspace{-4mm}
\end{figure}
For our proposed algorithms, Fig.~\ref{fig1} plots the optimal EE of hybrid arrays with different maximum PA efficiency $\eta_{\max}$ and dynamic circuit power coefficient $\epsilon$, where $M = 16$, $K = 16$, and $r_{\rm dl}=60$ Mbps. Coherent beamforming is shown to outperform its \mbox{non-coherent} counterpart due to the availability of explicit CSI. It is observed that EE under both coherent and \mbox{non-coherent} beamforming deteriorates as $\eta_{\max}$ decreases and/or $\epsilon$ increases. Coherent beamforming displays quicker decrease than \mbox{non-coherent} beamforming due to the fact that \mbox{non-coherent} beamforming has larger transmit powers and therefore is less sensitive to the circuit and PA consumptions. In addition, the improvement of EE, stemming from $\eta_{\max}$, is much higher than from $\epsilon$, indicating that getting more efficient PAs can be preferable to efficient circuit designs.

\begin{figure}[!t]
\centering
\subfloat[Optimal transmit duration]{\hspace{-4mm} \label{fig5}
\includegraphics[scale=0.28]{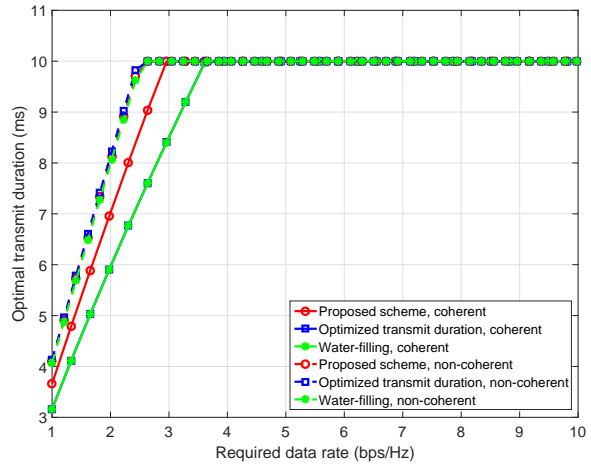}}
\vspace{-4mm}
\hfil
\centering
\subfloat[Optimal transmit power]{\hspace{-4mm} \label{fig6}
\includegraphics[scale=0.28]{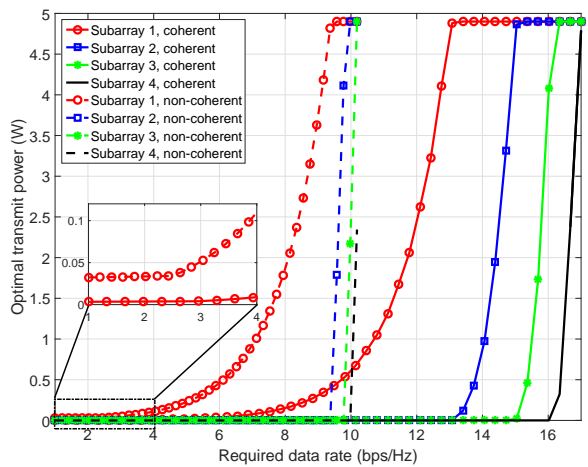}}
\caption{Optimal resource allocation results for the proposed algorithms considering both TPA and non-negligible circuit power ($M = 4$ and $K = 16$).}
\label{fig56}
\vspace{-4mm}
\end{figure}
Fig.~\ref{fig56} plots the optimal transmit power and duration under TPA and non-negligible circuit power, where a hybrid array with $4$ subarrays and $16$ antennas per subarray is considered. It is observed that the proposed algorithms are able to leverage the transmit powers and duration. Under both coherent and \mbox{non-coherent} beamforming, when $r_{\rm dl}$ is low, the transmit duration is less than $T$ and grows linearly with $r_{\rm dl}$. Meanwhile, the optimal transmit powers of the subarrays stay nearly unchanged. When $r_{\rm dl}$ is large, $T$ is used up for transmission, and the optimal transmit powers of the active subarrays increase exponentially to meet the growth of $r_{\rm dl}$.

Further, Fig.~\ref{fig5} shows that the optimal transmit duration of coherent beamforming is shorter than that of \mbox{non-coherent} beamforming. This is because coherent beamforming is superior in terms of SE and requires a shorter transmit time, thereby reducing the circuit power consumption and improving the EE. The optimal transmit durations of the proposed algorithms are larger than those of the benchmarks under coherent beamforming, since the benchmarks turn on all subarrays and therefore can finish transmission in a shorter time.

\begin{figure}[!t]
\centering
\subfloat[Optimal EE as function of time slot $T$]{\hspace{-4mm} \label{fig7} \includegraphics[scale=0.28]{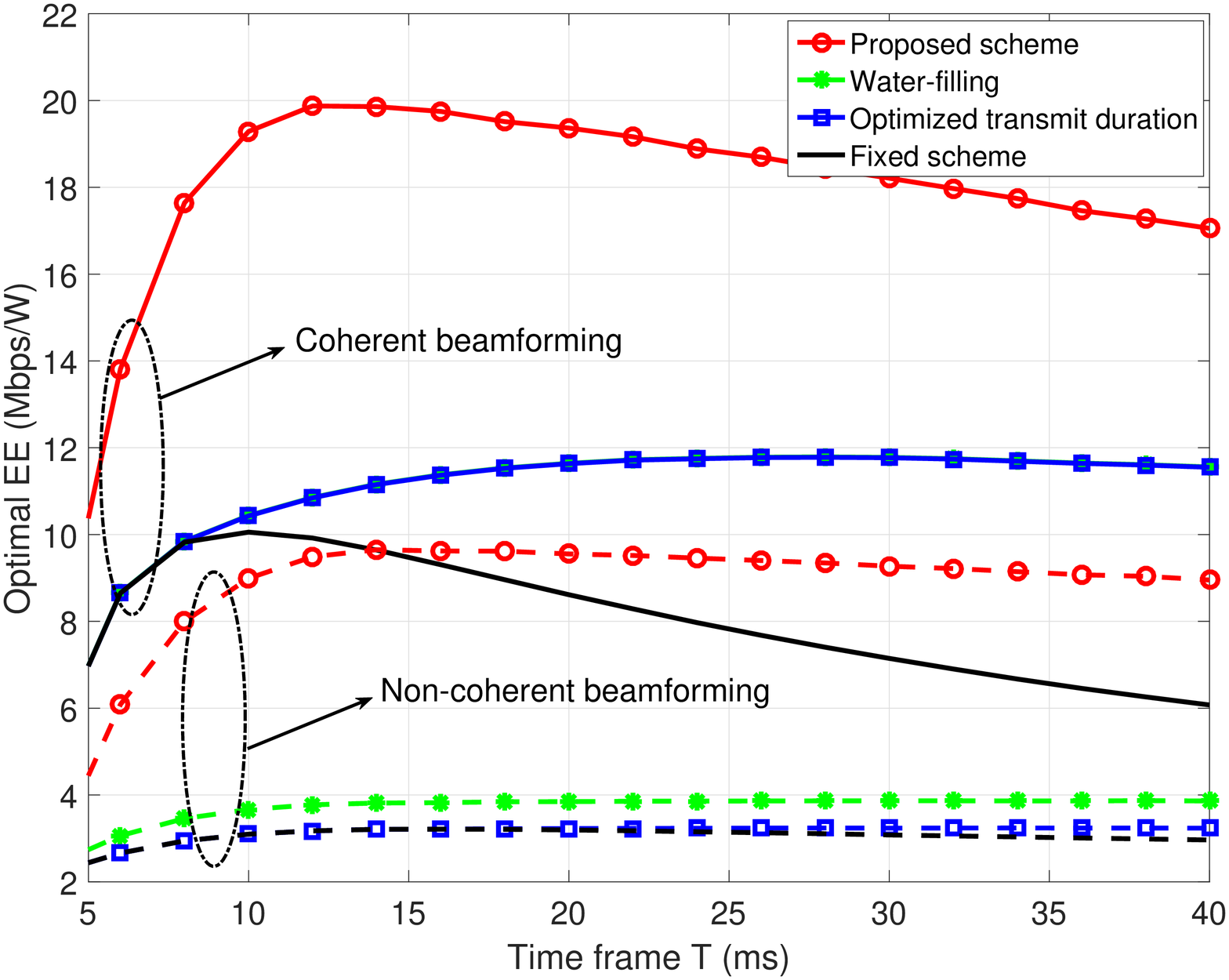}}
\vspace{-4mm}
\hfil
\centering
\subfloat[Optimal transmit duration as function of time slot $T$]{\hspace{-4mm} \label{fig8} \includegraphics[scale=0.28]{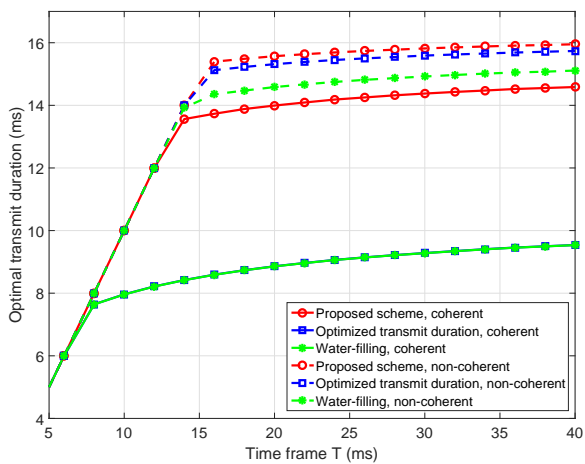}}
\caption{The effect of slot duration $T$ for different schemes of hybrid array with TPA and non-negligible circuit power ($M = 16$ and $K = 16$).}
\label{fig78}
\vspace{-4mm}
\end{figure}
Fig.~\ref{fig78} shows the impact of $T$ on the optimal EE and transmit duration, where the total data requirement is 400 kbits per $T$. The proposed algorithms can outperform the benchmarks in both coherent and \mbox{non-coherent} beamforming. Fig.~\ref{fig7} shows that the EE of the proposed algorithms first increases until $T=12$ ms, and then decreases. The reason is because when $T$ is small, $r_{\rm dl}$ needs to be large enough to support the total data requirement, and the hybrid array transmits throughout $T$, as shown in Fig.~\ref{fig8}. The total transmit power is high and dominates the EE. When $T<12$ ms, the EE improves as $T$ grows, since the transmit powers can decrease exponentially with the linear growth of $T$ and in turn the overall energy consumption decreases.

On the other hand, when $T>12$ ms, $T$ is excessively long and the subarrays only transmit for part of a slot, as shown in Fig.~\ref{fig8}. The circuit power consumption increases as $T$ grows, compromising the EE. This is particularly severer in the fixed scheme, which does not optimize the transmit duration and requires transmission across the entire slot of $T$. The EE curves of the optimized transmit duration scheme and water-filling scheme stay almost unchanged when $T>15$ ms. This is because the transmit powers of the two schemes are large and dominate over the circuit power consumption.
\section{Conclusion}\label{sec:8}
In this paper, the structure of the most energy-efficient transmit powers of all analog subarrays are discovered in hybrid arrays with \mbox{non-ideal} PAs and circuits, given a transmit duration. The structure, in turn, is able to fragment the \mbox{non-convex} feasible region of the transmit duration into disjoint segments with strict convexity or concavity. In both cases of coherent and \mbox{non-coherent} beamforming, our discovery enables the intractable \mbox{non-convex} maximization of EE under \mbox{non-ideal} PAs and non-negligible circuit power to be efficiently solved segment by segment with linear complexity. The optimality of the proposed approach is confirmed by significant gains in comparison with the state of the art.

\appendices
\section{Proof of~\eqref{pri_fact}}\label{Apx1}
\begin{IEEEproof}
Assume that, for the problem of interest, an optimal solution, denoted by Solution~I, does not satisfy~\eqref{pri_fact}. If any subarray $\pi(m)$, $m<m^*$, is inactive, activating subarray $\pi(m)$ and deactivating subarray $\pi(m^*)$ can provide an alternative solution to Solution~I, denoted by Solution~II, to the problem. Solution~II can be given by
\begin{equation}\nonumber
\left\{ {\begin{array}{*{20}{l}}
{0< {x_{\pi(i)}} \leq P_{\max}^2},&{i \in \left\{ {1,2,\cdots,{m^*}-1} \right\};}\\
{{x_{\pi(i)}} = 0},&{i \in \left\{ {{m^*},{m^*} + 1,\cdots,M} \right\},}
\end{array}} \right.
\end{equation}
which can also achieve the required data rate. According to~\eqref{Re_RateConstraint}, we attain ${x_{\pi(m)}}\kappa _{\pi(m)}^2={x_{\pi(m^*)}}\kappa _{\pi(m^*)}^2$ under \mbox{non-coherent} beamforming and $\sqrt {{x_{\pi(m)}}} {\kappa _{\pi(m)}} = \sqrt {{x_{\pi(m^*)}}} {\kappa _{\pi(m^*)}}$ under coherent beamforming, where $x_{\pi(m)} \leq x_{\pi(m^*)}$ since $\kappa_{\pi(m)} \geq \kappa_{\pi(m^*)}$.

According to~\eqref{EE_Opt_TPA}, we can obtain the difference of the total energy consumption between the two solutions, as given by
\begin{equation}\nonumber
\Delta E_{\rm{I,II,0}} = \left( {\sqrt{{x_{\pi(m)}}} - \sqrt{{x_{\pi(m^*)}}}} \right) t \leq 0,
\end{equation}
which indicates Solution~II is more energy-efficient.

Moreover, if any two subarrays $\pi(i)$ and $\pi(j)$, $i,j \leq m^*$, do not satisfy~\eqref{pri_fact}, i.e., $\kappa_{\pi(i)} \geq \kappa_{\pi(j)}$ and $x_{\pi(j)} > x_{\pi(i)}$, a possible solution for the EE maximization can be obtained by switching the roles of $x_{\pi(i)}$ and $x_{\pi(j)}$, which has the same energy consumption but a higher data rate, due to the higher received power at the receiver. This is because
\begin{equation}\nonumber
\left\{ \begin{array}{ll}
\hspace{-1mm} (x_{\pi(i)} \hspace{-0.5mm} - \hspace{-0.5mm} x_{\pi(j)})(\kappa_{\pi(j)}^2 \hspace{-0.5mm} - \hspace{-0.5mm} \kappa_{\pi(i)}^2) \geq 0, & {\text{non-coherent BF}};\\
\hspace{-1mm} (\sqrt{x_{\pi(i)}} \hspace{-0.5mm} - \hspace{-0.5mm} \sqrt{x_{\pi(j)}})(\kappa_{\pi(j)} \hspace{-0.5mm} - \hspace{-0.5mm} \kappa_{\pi(i)}) \geq 0, & {\text{coherent BF}}.
\end{array} \right.
\end{equation}
To this end, one can reduce $x_{\pi(i)}$ until the new solution achieves the target data rate, while still satisfying~\eqref{pri_fact}. The resultant solution consumes less energy and can be more energy-efficient. This concludes the proof of~\eqref{pri_fact}.
\end{IEEEproof}

\section{Proof of Theorem~\ref{PowerA_TPA}}\label{Apx2}
\begin{IEEEproof}
In the case that $M=1$, this theorem holds obviously. Therefore, this proof is focused on the case that $M \geq 2$. Assume that \eqref{G_EE_Opt} can have a solution, referred to as the first solution, satisfying~\eqref{pri_fact} but $x_{\pi(i)}$ is unnecessarily equal to $P_{\max}^2$ (as opposed to the theorem), as given by
\begin{equation}\nonumber
\left\{ {\begin{array}{*{20}{l}}
{0< {x_{\pi(i)}} \leq P_{\max}^2},&{i \in \left\{ {1,2,\cdots,{m_p}} \right\};}\\
{{x_{\pi(i)}} = 0},&{i \in \left\{ {{m_p} + 1,{m_p} + 2,\cdots,M} \right\}.}
\end{array}} \right.
\end{equation}
As a result, $m_p\geq m^*$.

With $0< m \leq {m_p-1}$, the second solution can be given by
\begin{equation}\nonumber
\{ \{x_{\pi(i)}\}_{{i \ne m,\ {m_p}}}, x_{\pi(m)}+\alpha, x_{\pi(m_p)}-\beta, t\},
\end{equation}
where $0<\alpha \leq \ x_{\pi(m-1)} - x_{\pi(m)}$ for $m>1$, $0<\alpha \leq \ P_{\max}^2 - x_{\pi(m)}$ for $m=1$ and $0<\beta \leq x_{\pi(m_p)}$, based on which~\eqref{pri_fact} still holds.

We proceed to prove that the second solution is more energy-efficient than the first in both cases of \mbox{non-coherent} and coherent beamforming. In the case of non-coherent beamforming, since the two solutions have the same data rate requirement ${r_{{\rm{dl}}}}$, according to~\eqref{Re_RateConstraint}, we have
\begin{equation}\nonumber
\begin{aligned}
& \sum\limits_{i \ne m}^{{m_p-1}} {{x_{\pi(i)}}\kappa _{\pi(i)}^2} \hspace{-1mm} + \hspace{-1mm} {x_{\pi(m)}}\kappa _{\pi(m)}^2 \hspace{-1mm} + \hspace{-1mm} {x_{\pi(m_p)}}\kappa _{\pi(m_p)}^2 = \\
& \sum\limits_{i \ne m}^{{m_p-1}} {{x_{\pi(i)}}\kappa _{\pi(i)}^2} \hspace{-1mm} + \hspace{-1mm} \left( {{x_{\pi(m)}} \hspace{-1mm} + \hspace{-1mm} \alpha } \right)\kappa _{\pi(m)}^2 \hspace{-1mm} + \hspace{-1mm} \left( {{x_{\pi(m_p)}} \hspace{-1mm} - \hspace{-1mm} \beta } \right)\kappa _{\pi(m_p)}^2.
\end{aligned}
\end{equation}
As a result, we attain $\alpha \kappa _{\pi(m)}^2 = \beta \kappa _{\pi(m_p)}^2$ and $0 < \alpha \leq \beta$ since $\kappa_{\pi(m)} \geq \kappa_{\pi(m_p)} \geq 0$. Given $0 < \alpha \leq \beta$ and $x_{\pi(m)} \geq x_{\pi(m_p)} \geq 0$, we have
\begin{equation}\nonumber
\alpha {x_{\pi(m_p)}} - \beta {x_{\pi(m)}} - \alpha \beta < 0,
\end{equation}
based on which, we have
\begin{equation}\nonumber
\begin{aligned}
\Delta E ^2 = &{\left( {\sqrt {{x_{\pi(m)}}} \hspace{-1mm} + \hspace{-1mm} \sqrt {{x_{\pi(m_p)}}} } \right)^2} \hspace{-1mm} - \hspace{-1mm} {\left( {\sqrt {{x_{\pi(m)}} \hspace{-1mm} + \hspace{-1mm} \alpha } \hspace{-1mm} + \hspace{-1mm} \sqrt {{x_{\pi(m_p)}} \hspace{-1mm} - \hspace{-1mm} \beta } } \right)^2} \\
= & \hspace{-1mm} - \hspace{-1mm} \alpha \hspace{-1mm} + \hspace{-1mm} \beta \hspace{-1mm} + \hspace{-1mm} 2 \sqrt {{x_{\pi(m)}}{x_{\pi(m_p)}}} \\
& \hspace{-1mm} - \hspace{-1mm} 2 \sqrt {{x_{\pi(m)}}{x_{\pi(m_p)}} \hspace{-1mm} + \hspace{-1mm} \alpha {x_{\pi(m_p)}} \hspace{-1mm} - \hspace{-1mm} \beta {x_{\pi(m)}} \hspace{-1mm} - \hspace{-1mm} \alpha \beta } > 0,
\end{aligned}
\end{equation}
and therefore
\begin{equation}\nonumber
{\left( {\sqrt {{x_{\pi(m)}}} \hspace{-1mm} + \hspace{-1mm} \sqrt {{x_{\pi(m_p)}}} } \right)}>{\left( {\sqrt {{x_{\pi(m)}} \hspace{-1mm} + \hspace{-1mm} \alpha } \hspace{-1mm} + \hspace{-1mm} \sqrt {{x_{\pi(m_p)}} \hspace{-1mm} - \hspace{-1mm} \beta } } \right)}>0.
\end{equation}

Given the TPA model and $t$, and according to~\eqref{EE_Opt_TPA}, we can have the difference of the total energy consumption between the two solutions, as given by
\begin{equation}\nonumber
\begin{aligned}
\Delta {E_{{\rm{I,II,1}}}}=& \left[ \left( {\sqrt {{x_{\pi(m)}}}  + \sqrt {{x_{\pi(m_p)}}}} \right) - \right. \\
&\left. \left( {\sqrt {{x_{\pi(m)}} + \alpha }  + \sqrt {{x_{\pi(m_p)}} - \beta } } \right) \right]t > 0,
\end{aligned}
\end{equation}
If $x_{\pi(m_p)}$ is reduced to zero, the ${\pi(m_p)}$-th subarray is in the idle mode and therefore consumes less energy. Therefore, the second solution is more energy-efficient than the first.

In the case of coherent beamforming, given $r_{\rm dl}$ and~\eqref{Re_RateConstraint}, we can have the following equality:
\begin{equation}\nonumber
\begin{aligned}
& \sum\limits_{i \ne m}^{{m_p-1}} \hspace{-1mm} {\sqrt {{x_{\pi(i)}}} {\kappa _{\pi(i)}}} \hspace{-1mm} + \hspace{-1mm} \sqrt {{x_{\pi(m)}}} {\kappa _{\pi(m)}} \hspace{-1mm} + \hspace{-1mm} \sqrt {{x_{\pi(m_p)}}} {\kappa _{\pi(m_p)}} = \\
& \sum\limits_{i \ne m}^{{{m_p}-1}} \hspace{-1mm} {\sqrt {{x_{\pi(i)}}} {\kappa _{\pi(i)}}} \hspace{-1mm} + \hspace{-1mm} \sqrt {{x_{\pi(m)}} \hspace{-1mm} + \hspace{-1mm} \alpha } {\kappa _{\pi(m)}} \hspace{-1mm} + \hspace{-1mm} \sqrt {{x_{\pi(m_p)}} \hspace{-1mm} - \hspace{-1mm} \beta } {\kappa _{\pi(m_p)}}.
\end{aligned}
\end{equation}

Define an auxiliary variable:
\begin{equation}\nonumber
\begin{aligned}
{\Gamma _{m,{m_p}}} & \buildrel \Delta \over = \sqrt {{x_{\pi(m)}}} {\kappa _{\pi(m)}} + \sqrt {{x_{\pi(m_p)}}} {\kappa _{\pi(m_p)}} \\
& = \sqrt {{x_{\pi(m)}} + \alpha } {\kappa _{\pi(m)}} + \sqrt {{x_{\pi(m_p)}} - \beta } {\kappa _{\pi(m_p)}}.
\end{aligned}
\end{equation}

With mathematic manipulation, we can have
\begin{equation}\label{Rel_l3_coh1}
\sqrt {{x_{\pi(m)}}} \hspace{-1mm} + \hspace{-1mm} \sqrt {{x_{\pi(m_p)}}} \hspace{-1mm} = \hspace{-1mm} \frac{{{\Gamma _{m,{m_p}}}}}{{{\kappa _{\pi(m)}}}} \hspace{-1mm} + \hspace{-1mm} \sqrt {{x_{\pi(m_p)}}} \left( {1 \hspace{-1mm} - \hspace{-1mm} \frac{{{\kappa _{\pi(m_p)}}}}{{{\kappa _{\pi(m)}}}}} \right),
\end{equation}
and
\begin{equation}\label{Rel_l3_coh2}
\sqrt {{x_{\pi(m)}} \hspace{-1mm} + \hspace{-1mm} \alpha } \hspace{-1mm} + \hspace{-1mm} \sqrt {{x_{\pi(m_p)}} \hspace{-1mm} - \hspace{-1mm} \beta } \hspace{-1mm} = \hspace{-1mm} \frac{{{\Gamma _{m,{m_p}}}}}{{{\kappa _{\pi(m)}}}} \hspace{-1mm} + \hspace{-1mm} \sqrt {{x_{\pi(m_p)}} \hspace{-1mm} - \hspace{-1mm} \beta } \hspace{-1mm} \left( \hspace{-1mm} {1 \hspace{-1mm} - \hspace{-1mm} \frac{{{\kappa _{\pi(m_p)}}}}{{{\kappa _{\pi(m)}}}}} \hspace{-1mm} \right) \hspace{-1mm} .
\end{equation}

By \eqref{Rel_l3_coh1} and \eqref{Rel_l3_coh2}, the difference of total energy under coherent beamforming, denoted by $\Delta {E_{{\rm{I,II,2}}}}$, can be given by
\begin{equation}\nonumber
\begin{aligned}
\Delta {E_{{\rm{I,II,2}}}} & \hspace{-1mm} = \hspace{-1mm} \left[ { \hspace{-1mm} \left( \hspace{-1mm} {\sqrt {{x_{\pi(m)}}} \hspace{-1mm} + \hspace{-1.5mm} \sqrt {{x_{\pi(m_p)}}} } \right) \hspace{-1mm} - \hspace{-1mm} \left( \hspace{-1mm} {\sqrt {{x_{\pi(m)}} \hspace{-1mm} + \hspace{-1mm} \alpha } \hspace{-1mm} + \hspace{-1mm} \sqrt {{x_{\pi(m_p)}} \hspace{-1mm} - \hspace{-1mm} \beta } } \hspace{-0.5mm} \right) \hspace{-0.5mm} } \right] \hspace{-1mm} t \\
& \hspace{-1mm} = \hspace{-1mm} \left( {\sqrt {{x_{\pi(m_p)}}} \hspace{-1mm} - \hspace{-1mm} \sqrt {{x_{\pi(m_p)}} \hspace{-1mm} - \hspace{-1mm} \beta } } \right)\left( {1 \hspace{-1mm} - \hspace{-1mm} \frac{{{\kappa _{\pi(m_p)}}}}{{{\kappa _{\pi(m)}}}}} \right)t,
\end{aligned}
\end{equation}
Apparently, $\Delta {E_{{\rm{I,II,2}}}}>0$, since $0<\beta \leq x_{\pi(m_p)}$, if ${\kappa _{\pi(m)} > \kappa _{\pi(m_p)}}>0$; and $\Delta {E_{{\rm{I,II,2}}}}=0$ if ${\kappa _{\pi(m)} = \kappa _{\pi(m_p)}}$. Note that if $x_{\pi(m_p)}$ is reduced to zero, the ${\pi(m_p)}$-th subarray is in the idle mode and therefore consumes less energy. As a result, the second solution is more energy-efficient than the first solution under coherent beamforming.

In light of this analysis, we can pump up the transmit powers of the subarrays one-by-one from the left end of~\eqref{pri_fact}, while reducing the non-zero transmit powers of the subarrays from the right side. Without violating $r_{\rm dl}$, the EE of the hybrid array can monotonically increase until the transmit powers of the subarrays satisfy $x_{\pi(1)}=x_{\pi(2)}=\cdots=x_{\pi(m^*-1)} =P_{\max}^2 \geq x_{\pi(m^*)}>x_{\pi(m^*+1)}=\cdots =x_{\pi(M)}=0$.

It is obvious that if subarray ${\pi(m^*)}$ is in the idle mode, the rate requirement cannot be met, as given by
\begin{equation}\label{Bset_m_f}
\left\{ {\begin{array}{*{20}{l}}
{\sum\limits_{m = 1}^{{m^*} - 1} {P_{\max}^2\kappa _{\pi(m)}^2} < \theta \left( t \right)},&{{\text{non-coherent BF}}};\\
{\sum\limits_{m = 1}^{{m^*} - 1} {{P_{\max}}{\kappa _{\pi(m)}}} < \sqrt {\theta \left( t \right)}},&{{\text{coherent BF}}}.
\end{array}} \right.
\end{equation}
If subarray ${\pi(m^*)}$ transmits with $P_{\max}$, the achievable data rate would exceed $r_{\rm dl}$, as given by
\begin{equation}\label{Bset_m_b}
\left\{ {\begin{array}{*{20}{l}}
\hspace{-1mm} {\theta \left( t \right) \hspace{-1mm} < \hspace{-1mm} \sum\limits_{m = 1}^{{m^*} - 1} \hspace{-2mm} {P_{\max}^2\kappa _{\pi(m)}^2} \hspace{-1mm} + \hspace{-1mm} {P_{\max}^2\kappa _{\pi(m^*)}^2}},\text{non-coherent BF};\\
\hspace{-1mm} {\sqrt {\theta \left( t \right)} \hspace{-1mm} < \hspace{-1mm} \sum\limits_{m = 1}^{{m^*} - 1} \hspace{-2mm} {{P_{\max}}{\kappa _{\pi(m)}}} \hspace{-1mm} + \hspace{-1mm} {{P_{\max}}{\kappa _{\pi(m^*)}}}},\text{coherent BF}.
\end{array}} \right.
\end{equation}

By combining~\eqref{Bset_m_f} and~\eqref{Bset_m_b}, $m^*$ can be identified, as specified in~\eqref{Bset_m}. According to~\eqref{Re_RateConstraint}, the optimal transmit powers of the $M$ analog subarrays can be given by \eqref{Pi_TPA}.
\end{IEEEproof}

\end{document}